\PassOptionsToPackage{unicode}{hyperref}
\PassOptionsToPackage{hyphens}{url}
\documentclass[
]{article}
\usepackage{lmodern}
\usepackage{amssymb,amsmath}
\usepackage{ifxetex,ifluatex}
\ifnum 0\ifxetex 1\fi\ifluatex 1\fi=0 
  \usepackage[T1]{fontenc}
  \usepackage[utf8]{inputenc}
  \usepackage{textcomp} 
\else 
  \usepackage{unicode-math}
  \defaultfontfeatures{Scale=MatchLowercase}
  \defaultfontfeatures[\rmfamily]{Ligatures=TeX,Scale=1}
\fi
\IfFileExists{upquote.sty}{\usepackage{upquote}}{}
\IfFileExists{microtype.sty}{
  \usepackage[]{microtype}
  \UseMicrotypeSet[protrusion]{basicmath} 
}{}
\makeatletter
\@ifundefined{KOMAClassName}{
  \IfFileExists{parskip.sty}{%
    \usepackage{parskip}
  }{
    \setlength{\parindent}{0pt}
    \setlength{\parskip}{6pt plus 2pt minus 1pt}}
}{
  \KOMAoptions{parskip=half}}
\makeatother
\usepackage{xcolor}
\IfFileExists{xurl.sty}{\usepackage{xurl}}{} 
\IfFileExists{bookmark.sty}{\usepackage{bookmark}}{\usepackage{hyperref}}
\hypersetup{
  pdftitle={Accelerating Quantum Materials Characterization: Hybrid Active Learning for Autonomous Spin Wave Spectroscopy},
  hidelinks,
  pdfcreator={LaTeX via pandoc}}
\usepackage{longtable,booktabs}
\usepackage{etoolbox}
\makeatletter
\patchcmd\longtable{\par}{\if@noskipsec\mbox{}\fi\par}{}{}
\makeatother
\IfFileExists{footnotehyper.sty}{\usepackage{footnotehyper}}{\usepackage{footnote}}
\makesavenoteenv{longtable}
\usepackage{graphicx}
\makeatletter
\def\maxwidth{\ifdim\Gin@nat@width>\linewidth\linewidth\else\Gin@nat@width\fi}
\def\maxheight{\ifdim\Gin@nat@height>\textheight\textheight\else\Gin@nat@height\fi}
\makeatother
\setkeys{Gin}{width=\maxwidth,height=\maxheight,keepaspectratio}
\makeatletter
\def\fps@figure{htbp}
\makeatother
\providecommand{\tightlist}{%
  \setlength{\itemsep}{0pt}\setlength{\parskip}{0pt}}
\newlength{\cslhangindent}
\newenvironment{cslreferences}%
  {}%
  {\par}

\title{Accelerating Quantum Materials Characterization: Hybrid Active
Learning for Autonomous Spin Wave Spectroscopy}
\author{}
\date{}

\begin{document}
\maketitle

\textbf{Authors:} William Ratcliff II,¹*

¹NIST Center for Neutron Research, National Institute of Standards and
Technology, Gaithersburg, MD 20899, USA

*Corresponding author: william.ratcliff@nist.gov

\begin{center}\rule{0.5\linewidth}{0.5pt}\end{center}

\hypertarget{abstract}{%
\subsection{Abstract}\label{abstract}}

Autonomous neutron spectroscopy must solve three distinct tasks:
\textbf{detection} (where is the signal?), \textbf{inference} (which
Hamiltonian governs it?), and \textbf{refinement} (what are the
parameters?). No single controller solves all three equally well. We
present TAS-AI, a \textbf{hybrid agnostic→physics-informed} framework
for autonomous triple-axis spin-wave spectroscopy that separates these
tasks explicitly. In blind reconstruction benchmarks, where the
controller must map an unknown response surface without prior model
structure, model-agnostic methods such as random sampling, coarse grids,
and Gaussian-process mappers reach a global error threshold more
reliably and with fewer measurements than physics-informed planning.
This supports the central claim that discovery and inference are
distinct tasks requiring distinct controllers. Once signal structure is
localized, the physics-informed stage performs \textbf{in-loop
Hamiltonian discrimination and parameter refinement}: in a controlled
square-lattice test between nearest-neighbor-only and \(J_1\)-\(J_2\)
Hamiltonians, TAS-AI reaches a decisive AIC-derived evidence ratio
(\textgreater100) in fewer than 10 measurements, while motion-aware
scheduling cuts wall-clock time by 32\% at a fixed measurement budget.
We further identify a failure mode of posterior-weighted design,
\textbf{algorithmic myopia}, in which the planner over-refines the
current leading model while under-sampling low-intensity falsification
probes. A \textbf{constrained falsification channel} sharply reduces
time spent committed to the wrong model and accelerates correct model
selection without modifying the Bayesian inference engine. In controlled
two-model ablations, both a simple deterministic top-two
max-disagreement rule and an LLM-based audit committee achieve this gain
under identical constraints, showing that the active ingredient is the
falsification principle rather than the specific implementation. In a
targeted multi-model stress test, however, top-two max-disagreement
becomes structurally blind because the decisive falsifier separates the
current leader from a lower-ranked model rather than from the runner-up;
a broader falsification policy and the LLM both recover. The LLM
additionally handles diverse ambiguity descriptions through the same
interface without per-problem engineering. We demonstrate the full
workflow \emph{in silico} using a high-fidelity digital twin and provide
an open-source Python implementation.

\textbf{Keywords:} autonomous experiments, neutron scattering, spin
waves, Bayesian optimization, active learning, triple-axis spectrometer,
strategic falsification

\begin{center}\rule{0.5\linewidth}{0.5pt}\end{center}

\hypertarget{introduction}{%
\subsection{1. Introduction}\label{introduction}}

Spin-wave spectroscopy is a gateway to quantum materials. It reveals
exchange pathways, anisotropies, and emergent excitations in ordered
magnets, multiferroics, and unconventional superconductors that underpin
functional behavior.\textsuperscript{1--6} Triple-axis spectrometers
remain the instrument of choice for detailed spin-wave
characterization,\textsuperscript{7} but even modern multiplexed
variants survey narrow regions of reciprocal space at a time. Their
point-by-point measurement paradigm makes efficient use of beam time a
first-order concern. A practical autonomous controller for these
instruments must decide what to measure next, how long to count, and
when to stop.

The key claim of this paper is that autonomous spectroscopy is
\textbf{not one task}. It is a sequence of distinct tasks with different
objectives:

\begin{enumerate}
\def\labelenumi{\arabic{enumi}.}
\tightlist
\item
  \textbf{Detection / coarse reconstruction:} where in \((Q,E)\) space
  does signal exist?
\item
  \textbf{Model inference:} which Hamiltonian family best explains the
  observed response?
\item
  \textbf{Parameter refinement:} what are the posterior distributions of
  the physical parameters within the winning model?
\end{enumerate}

These tasks reward different acquisition strategies. Global mapping of
an unknown response surface is naturally aligned with
\textbf{model-agnostic} methods such as Gaussian-process-based active
learning.\textsuperscript{8--15} Once a plausible model family and
signal support exist, however, the objective changes: the experiment
should spend time where the forward model is most informative about
parameters or where competing Hamiltonians diverge most strongly. That
second regime favors \textbf{physics-informed} experimental design.

This distinction matters for neutron spectroscopy. Current autonomous
neutron methods are largely model-agnostic. gpCAM-style workflows and
related active-learning strategies efficiently map scattering landscapes
but treat the response surface as a black box.\textsuperscript{16,17}
The Log-GP approach of Teixeira Parente \emph{et al.} showed that sparse
neutron spectroscopy can be reconstructed effectively in log-intensity
space, preserving physical non-negativity while remaining
model-free.\textsuperscript{18} In neutron reflectometry, AutoRefl
applies active learning to drive measurement schedules that distinguish
competing structural models in real time.\textsuperscript{19} On a
related neutron platform, autonomous control has also been demonstrated
for spin-echo response-function data collection, where the controller
decides which Fourier times to sample under shared time and motion
budgets.\textsuperscript{20} A complementary physics-informed system,
ANDiE, demonstrated autonomous neutron diffraction using a fixed
Weiss-model controller, but performs hypothesis testing only after the
autonomous data set is complete.\textsuperscript{21} These advances
motivate a sharper question: how should an autonomous spectrometer
operate when the \textbf{location of the signal is initially unknown}
and the \textbf{Hamiltonian itself is uncertain}?

Here we argue that no single controller should be asked to solve all
stages simultaneously. Instead, we introduce \textbf{TAS-AI}, a hybrid
workflow that begins with agnostic discovery and then hands control to a
physics-informed, motion-aware inference engine once enough structure
exists to justify model-based planning. This framing turns an apparent
weakness into a design principle: if agnostic methods outperform
physics-informed planning on blind global reconstruction, that is
evidence that discovery and inference are distinct problems rather than
evidence against model-based autonomy.

A second challenge arises even after the system enters the
physics-informed regime. Posterior-weighted acquisition functions are
powerful, but they can become \textbf{self-reinforcing}. If an incorrect
model takes an early lead, the planner may repeatedly choose
measurements that refine that model rather than measurements that could
falsify it. In our closed-loop gapped-versus-gapless tests, this
manifests as a silent-data failure mode: low-intensity gap-sensitive
regions produce near-zero counts for multiple hypotheses, so greedy
posterior-weighted planning delays the very probes that would break the
tie. We call this vulnerability \textbf{algorithmic myopia}. In this
manuscript we show how physics-aware contrast policies and forced
coverage mitigate it, and we describe a constrained strategic audit
layer---implemented here as an LLM overseer pilot---that can request a
small number of falsification probes without replacing the numerical
inference engine.

The paper makes four contributions.

\begin{enumerate}
\def\labelenumi{\arabic{enumi}.}
\tightlist
\item
  \textbf{Task-based hybrid autonomy.} We define autonomous TAS as a
  sequence of detection, inference, and refinement problems, and show
  that a hybrid agnostic→physics workflow is the appropriate operating
  regime.
\item
  \textbf{In-loop model discrimination and parameter refinement.} TAS-AI
  maintains posterior weights over competing Hamiltonians and uses those
  weights to select measurements that are simultaneously informative and
  discriminative.
\item
  \textbf{Time-aware physical planning.} The acquisition function
  explicitly accounts for motor motion costs, and an optional Monte
  Carlo Tree Search (MCTS) planner reduces path inefficiency in batched
  trajectories.
\item
  \textbf{Constrained falsification channels for algorithmic myopia.} We
  identify silent-data posterior lock-in as a real failure mode and show
  that a bounded falsification channel---whether instantiated as a
  simple max-disagreement rule or as an LLM committee---eliminates
  wrong-leader dwell and accelerates decisive recovery under identical
  constraints. A targeted multi-model stress test shows where top-two
  heuristics become structurally blind. The LLM implementation offers
  generality across diverse ambiguity descriptions without per-problem
  engineering---an architectural advantage for a system designed to
  handle unknown Hamiltonians.
\end{enumerate}

The manuscript is organized around those tasks. Section 2 introduces the
hybrid workflow and formal problem statement. Section 3 describes the
agnostic discovery policy, the physics-informed inference engine, the
instrument model, and batch planning. Section 4 presents the benchmark
results in the order the workflow is actually used: blind discovery,
physics-informed inference, wall-clock optimization, and a fully
integrated hybrid handoff. Section 5 addresses the algorithmic myopia
failure mode and presents the constrained audit layer, including its
design, pilot closed-loop results, and targeted ablations. Section 6
discusses outlook topics including hypothesis generation and hardware
deployment. Section 7 concludes.

\begin{center}\rule{0.5\linewidth}{0.5pt}\end{center}

\hypertarget{hybrid-workflow-and-problem-statement}{%
\subsection{2. Hybrid workflow and problem
statement}\label{hybrid-workflow-and-problem-statement}}

\hypertarget{three-experimental-tasks-and-one-hybrid-controller}{%
\subsubsection{2.1 Three experimental tasks and one hybrid
controller}\label{three-experimental-tasks-and-one-hybrid-controller}}

Figure 1 summarizes the TAS-AI workflow. The controller is divided into
four layers.

\begin{enumerate}
\def\labelenumi{\arabic{enumi}.}
\tightlist
\item
  \textbf{Agnostic discovery.} A Log-GP mapper performs initial coverage
  of accessible \((Q,E)\) space and proposes additional measurements to
  localize signal support.
\item
  \textbf{Physics-informed discrimination and refinement.} Once signal
  structure is present, a Hamiltonian-aware planner ranks candidate
  measurements by expected information gain and model contrast per unit
  time.
\item
  \textbf{Motion-aware sequencing.} The queue is ordered to reduce
  wall-clock overhead, either greedily or with a short-horizon MCTS
  batch planner when path dependence matters.
\item
  \textbf{Strategic audit layer (optional).} When posterior-weighted
  planning becomes myopic, a constrained router may request a small
  number of falsification probes before normal planning resumes.
\end{enumerate}

This decomposition is deliberate. It allows the experiment to use the
right inductive bias at the right time: broad exploration when little is
known, then model-based exploitation once the physics becomes
informative. This staged approach mirrors how experienced
experimentalists typically organize beam-time campaigns.

\begin{figure}
\centering
\includegraphics[width=6in,height=\textheight]{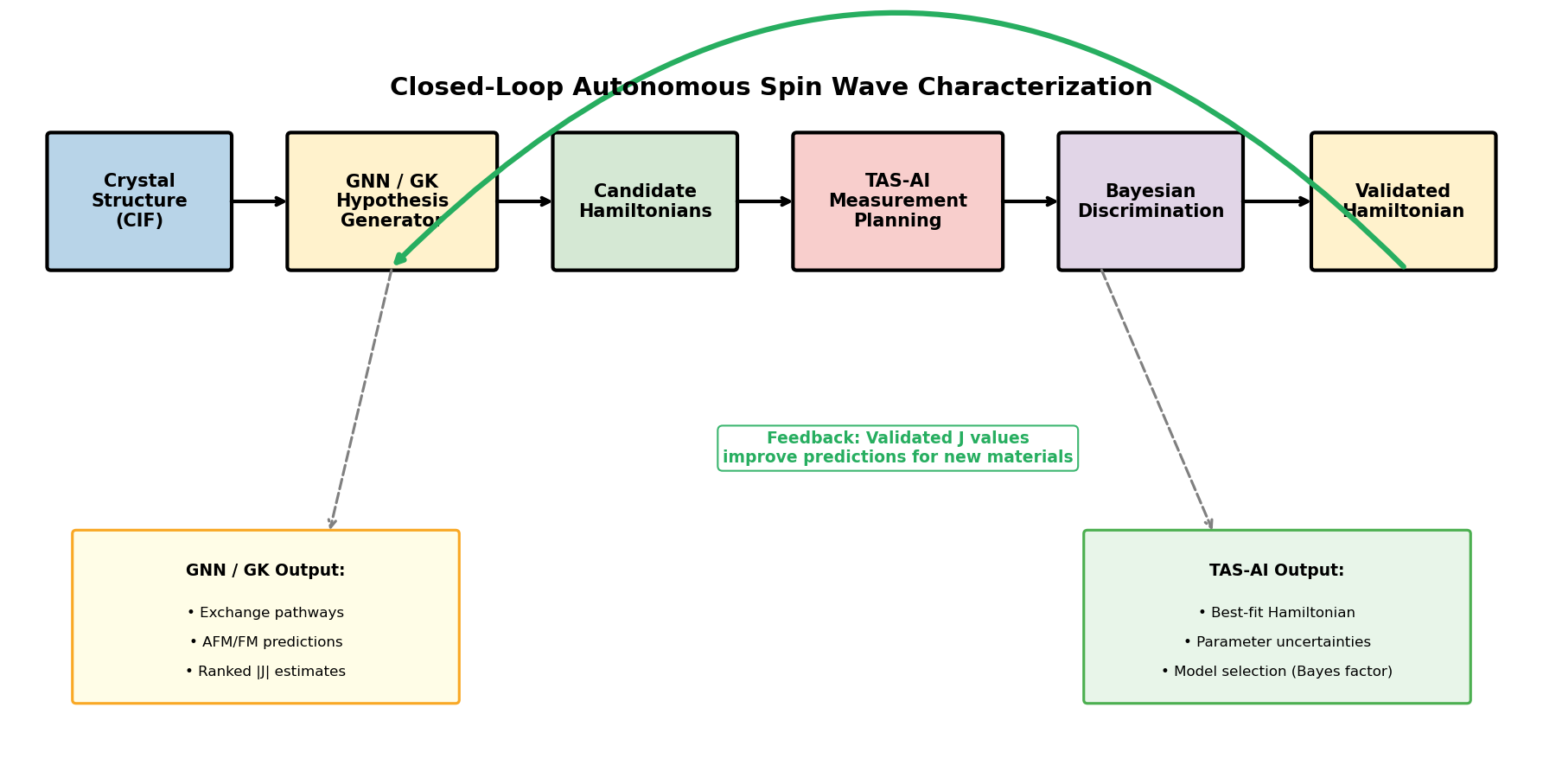}
\caption{Hybrid TAS-AI workflow. The controller begins with agnostic Log-GP mapping to localize signal, hands off to physics-informed discrimination/refinement once structure is detected, and uses motion-aware sequencing to optimize wall-clock time. An optional constrained audit layer can request targeted falsification probes under the same kinematic and safety constraints enforced by the numerical planners.}
\end{figure}

\hypertarget{sequential-decision-problem}{%
\subsubsection{2.2 Sequential decision
problem}\label{sequential-decision-problem}}

Consider a magnetic material with an unknown Hamiltonian model \(M\) and
unknown parameter vector \(\theta\) (e.g., \(J_1\), \(J_2\), and \(D\)).
The autonomous experiment collects a sequence of measurements

\[
\mathcal{D}_N = \{(Q_i,E_i,I_i,\sigma_i,t_i)\}_{i=1}^{N},
\]

where \(Q_i\) and \(E_i\) are the measured momentum-energy coordinates,
\(I_i\) is the observed intensity, \(\sigma_i\) is the corresponding
uncertainty estimate, and \(t_i\) is the counting time. The controller
must infer both the discrete model identity and the continuous parameter
posterior while minimizing total wall-clock time.

The controller therefore seeks a policy

\[
\pi : \mathcal{D}_N \mapsto (Q_{N+1},E_{N+1},t_{N+1})
\]

that maps the current data set to the next momentum-energy point and
count time. In the hybrid workflow this policy is not fixed: discovery
and inference are handled by different controllers because the utility
functions are different.

\begin{center}\rule{0.5\linewidth}{0.5pt}\end{center}

\hypertarget{methods}{%
\subsection{3. Methods}\label{methods}}

\hypertarget{agnostic-discovery-with-enhanced-log-gp}{%
\subsubsection{3.1 Agnostic discovery with enhanced
Log-GP}\label{agnostic-discovery-with-enhanced-log-gp}}

The agnostic phase uses Gaussian-process regression in
\textbf{log-intensity space},

\[
z(Q,E)=\log[I(Q,E)+\epsilon],
\]

where \(I(Q,E)\) is the scattering intensity and \(\epsilon\) is a small
positive offset that regularizes the transform in low-count regions.
Following the neutron-spectroscopy reconstruction framework of Teixeira
Parente \emph{et al.}\textsuperscript{18,22}, this guarantees
non-negative predictions after exponentiation and is well suited to
sparse discovery of unknown signal support.

In practice, raw log-space variance maximization exhibits a boundary
failure mode: GP variance peaks at the edges of a bounded domain, and
log-variance treats dim background regions as equally valuable as bright
signal regions. For the agnostic discovery phase, we therefore wrap the
base Log-GP regression in four additional safeguards to produce what we
call the \textbf{enhanced Log-GP} discovery policy:

\begin{enumerate}
\def\labelenumi{\arabic{enumi}.}
\tightlist
\item
  a small coarse initialization grid,
\item
  linear-intensity variance weighting for acquisition,
\item
  consumed-area exclusion using resolution-sized ellipses, and
\item
  a soft one-dimensional energy taper to suppress edge-locking.
\end{enumerate}

These modifications leave the method model agnostic while making it
behave like a realistic neutron-mapping policy rather than an
unconstrained variance maximizer. In the current implementation,
candidate scoring uses linear-space variance proxies exposed by the
surrogate backend; when log-space mean and variance are directly
available, this can be written as the usual log-normal conversion
detailed in Supplementary Note S1, where the taper is also described.

Throughout the paper we distinguish this \textbf{enhanced Log-GP} policy
(used in the full hybrid workflow of Figures 9 and 10 and in the
blind-reconstruction benchmarks of Table 1) from a \textbf{bare Log-GP}
variant that uses only the underlying log-space GP regression without
the taper, clamp, or consumed-region safeguards. The bare variant
appears in the audit ablation benchmarks (Tables 4 and 5), where a
simpler explorer suffices to test the lock-in failure mode in a
controlled setting.

\hypertarget{physics-informed-acquisition-for-parameter-refinement}{%
\subsubsection{3.2 Physics-informed acquisition for parameter
refinement}\label{physics-informed-acquisition-for-parameter-refinement}}

The handoff from agnostic discovery to physics-informed inference is
triggered automatically once the discovery stage has identified a
localized candidate signal region sufficient to instantiate the physics
model on a restricted action set; in the present implementation this is
a heuristic rule based on the seeded coarse survey plus the first active
Log-GP localization step, after which the controller switches to
motion-aware local exploration and Hamiltonian-aware refinement. The
acquisition function ranks candidate measurements by expected
information gain per unit wall-clock time:

\[
\alpha(Q,E)=\frac{[\mathrm{InfoGain}(Q,E)]^\gamma}{\mathrm{CountTime}(Q,E)+\mathrm{MoveTime}(Q,E)},
\]

where \(\alpha(Q,E)\) is the acquisition score,
\(\mathrm{InfoGain}(Q,E)\) is the expected local information gain at the
candidate point, \(\mathrm{CountTime}(Q,E)\) is the estimated counting
time, and \(\mathrm{MoveTime}(Q,E)\) is the corresponding motor-motion
overhead. The exploration exponent \(\gamma = 0.7\) is used throughout
as a stable default within the tested range
\(\gamma \in \{0.5, 0.7, 0.9\}\) (Supplementary Note S3.2).

Inside the real-time loop, the posterior is approximated locally as
Gaussian. For covariance matrix \(\Sigma\) and measurement uncertainty
\(\sigma\), the expected information gain is approximated by

\[
\mathrm{InfoGain}\approx \frac{1}{2}\log\!\left[1+\frac{1}{\sigma^2}
\left(\frac{\partial S}{\partial \theta}\right)^\mathsf{T}
\Sigma
\left(\frac{\partial S}{\partial \theta}\right)\right].
\]

Here \(S(Q,E;\theta)\) denotes the forward-model intensity at the
candidate point, \(\partial S/\partial\theta\) is its parameter
gradient, \(\Sigma\) is the local posterior covariance of the parameter
vector \(\theta\), and \(\sigma\) is the predicted observation
uncertainty at that point. This Laplace-style approximation is fast
enough for in-loop ranking of many candidate points. More expensive MCMC
updates are reserved for batch boundaries or offline validation when
local diagnostics fail. We note that the resulting uncertainty intervals
are useful as fast local surrogates for measurement ranking and routing,
but they are not calibrated Bayesian credible intervals: empirical
coverage in the refinement benchmark is substantially below nominal
(Supplementary Note S3.4).

\hypertarget{in-loop-model-discrimination}{%
\subsubsection{3.3 In-loop model
discrimination}\label{in-loop-model-discrimination}}

To discriminate among competing Hamiltonians \(M_k\), TAS-AI maintains
model scores

\[
P(M_k\mid \mathcal D) \propto P(\mathcal D\mid M_k)\,P(M_k).
\]

Here \(\mathcal D\) denotes the current accumulated data,
\(P(\mathcal D\mid M_k)\) is the model evidence term for candidate
Hamiltonian \(M_k\), and \(P(M_k)\) is its prior weight. Because
rigorous marginal likelihood estimation is too expensive for a
sub-second loop, we use \textbf{AIC-weight ratios} as a pragmatic
real-time proxy:

\[
w_k \propto \exp\!\left(-\frac{\mathrm{AIC}_k}{2}\right),
\qquad
R_{ij}^{\mathrm{AIC}}=\frac{w_i}{w_j}.
\]

Here \(w_k\) is the normalized relative weight of model \(M_k\), and
\(R_{ij}^{\mathrm{AIC}}\) is the evidence ratio comparing models \(M_i\)
and \(M_j\). The AIC itself is approximated as

\[
\mathrm{AIC}_k = \chi^2_k + 2p_k,
\]

where \(\chi^2_k\) is the weighted least-squares misfit of model \(M_k\)
to the current data and \(p_k\) is the number of free parameters in that
model.\textsuperscript{23} We originally tested BIC, but in sequential
settings the increasing sample count changes the BIC penalty term
\(p_k \ln n\) even when newly added measurements mostly reinforce
already identified features rather than opening a genuinely new
model-selection regime.\textsuperscript{24} The issue is not that
\(\chi^2\) stops dominating the absolute scale, but that the relative
complexity penalty drifts with \(n\) in a way that is awkward for a
real-time controller comparing models after each small batch. AIC keeps
the complexity penalty fixed and therefore behaves more predictably as
an engineering proxy for in-loop discrimination. Throughout the paper,
``decisive'' evidence refers to \textbf{AIC-derived evidence ratios}
exceeding 100, using the Kass--Raftery Bayes-factor scale only as a
heuristic reference rather than as a claim of formal marginal-likelihood
evaluation.\textsuperscript{25}

Chemically informed priors are motivated by Goodenough--Kanamori
heuristics. In the square-lattice test case, the prior weights are
initialized as \([0.10,0.10,0.10,0.70]\) for
\(\{\mathrm{NN},\mathrm{NN{+}D},J_1\!-\!J_2,\mathrm{Full}\}\). This
prior accelerates discrimination but also shapes it; a sensitivity
analysis appears in §5.4 and Supplementary Note S5.4.

During the discrimination phase, nuisance background terms are frozen at
values estimated during discovery. This prevents gapless hypotheses from
absorbing weak gap signal into floating offsets and artificially
improving their AIC scores.

\hypertarget{algorithmic-myopia-and-silent-data-failure}{%
\subsubsection{3.4 Algorithmic myopia and silent-data
failure}\label{algorithmic-myopia-and-silent-data-failure}}

Posterior-weighted acquisition can fail even when the underlying physics
model is expressive enough. The problem is not numerical instability but
\textbf{algorithmic myopia}. If an incorrect model takes an early lead,
the planner may keep selecting high-intensity points that refine that
model rather than low-intensity points that could overturn it.

In our gapped-versus-gapless closed-loop tests, this occurs when the
queue dwells inside the expected gap. Both the gapped model and gapless
competitors predict near-zero intensity there within experimental error,
so the measurements contribute little discriminative information.
Because those silent data points are also cheap to fit under simpler
models, AIC-weighted planning can delay the very gap-sensitive probes
needed to falsify the wrong hypothesis. We refer to this failure mode as
\textbf{silent-data posterior lock-in}. Here, by \textbf{silent data} we
mean measurements whose observed intensity is near zero across multiple
competing hypotheses and therefore carries little model-discriminative
information. As formally derived in Supplementary Note S4, the local
Laplace approximation of parameter information systematically suppresses
the cross-model log-likelihood ratio once a false leader dominates the
posterior, so the refinement utility on the bright branch overwhelms the
falsification value of an under-sampled weak feature even when the
latter is kinematically accessible.

Two mitigations are used in the numerical loop itself:

\begin{enumerate}
\def\labelenumi{\arabic{enumi}.}
\tightlist
\item
  \textbf{contrast-aware selection}, which explicitly rewards regions
  where competing models separate, and\\
\item
  \textbf{forced diagnostic coverage}, such as high-symmetry seed
  points, to ensure the queue contains early falsification probes.
\end{enumerate}

Section 5 describes a more flexible audit layer that operates on top of
these numerical mitigations.

\hypertarget{instrument-model-resolution-and-uncertainty}{%
\subsubsection{3.5 Instrument model, resolution, and
uncertainty}\label{instrument-model-resolution-and-uncertainty}}

All simulations use realistic thermal-neutron TAS parameters
representative of modern triple-axis spectrometers. We assume fixed
final energy \(E_f=14.7\) meV, realistic horizontal and vertical
collimations, PG(002) monochromator/analyzer optics, and standard
kinematic accessibility constraints.

Energy broadening is computed from the Cooper--Nathans resolution
formalism.\textsuperscript{26,27} At each candidate point \((Q,E)\) we
evaluate the resolution matrix, extract the energy full width at half
maximum, convert it to Gaussian width \(\sigma_E\), and broaden the
model in energy rather than performing a full four-dimensional
convolution. This approximation captures the dominant effect relevant to
point-by-point planning while remaining computationally lightweight. On
steep dispersive branches, however, neglecting the full \(Q\)-\(E\)
coupling can modestly underestimate apparent linewidths relative to a
full 4D resolution treatment.

Synthetic data combine Poisson counting noise with a conservative 3\%
systematic floor,

\[
\sigma_{\mathrm{sys}} = 0.03\,I + 10^{-4},
\]

where \(I\) is the predicted or observed local intensity in the
simulator. This floor prevents extremely bright points from dominating
the likelihood because of minor resolution mismatches.

\hypertarget{spin-wave-models-backends-and-benchmark-provenance}{%
\subsubsection{3.6 Spin-wave models, backends, and benchmark
provenance}\label{spin-wave-models-backends-and-benchmark-provenance}}

For the analytic benchmarks we use a square-lattice ferromagnet with
nearest-neighbor exchange \(J_1\), next-nearest-neighbor exchange
\(J_2\), and easy-axis anisotropy \(D\),

\[
\mathcal H = -J_1\sum_{\langle i,j\rangle}\mathbf S_i\cdot \mathbf S_j
            -J_2\sum_{\langle\langle i,j\rangle\rangle}\mathbf S_i\cdot \mathbf S_j
            -D\sum_i (S_i^z)^2.
\]

Here \(\mathbf S_i\) is the spin operator on site \(i\),
\(\langle i,j\rangle\) denotes nearest-neighbor pairs,
\(\langle\langle i,j\rangle\rangle\) next-nearest-neighbor pairs,
\(J_1\) and \(J_2\) are the exchange constants, and \(D\) is the
single-ion anisotropy. With the sign convention above, positive \(J_1\)
and \(J_2\) correspond to ferromagnetic exchange because the Hamiltonian
carries explicit leading minus signs. Along the measured \([H,H,0]\)
cut, the one-magnon branch is

\[
\omega(H) = 2S\left[2J_1(1-\cos 2\pi H) + 2J_2(1-\cos^2 2\pi H)\right] + D(2S-1).
\]

where \(S\) is the spin quantum number and \(H\) is the reduced
reciprocal-lattice coordinate along the scan. The dynamical structure
factor is modeled as a Lorentzian centered on the dispersion,

\[
S(Q,E)=\frac{A\eta}{(E-\omega(Q))^2+\eta^2}\,n(E,T)+\mathrm{background},
\]

where \(A\) is an overall intensity scale, \(\eta\) is the linewidth,
\(n(E,T)\) is the thermal occupation factor, and \texttt{background}
denotes an additive offset term. This form is fast enough for real-time
use in the analytic loop.

The closed-loop audit-layer pilots use a separate square-lattice
antiferromagnetic \(J_1\)-\(J_2\)-\(D\) model referenced to the ordering
vector \(\mathbf Q_{\mathrm{AF}}=(0.5,0.5,0)\):

\[
\omega(\mathbf q)=\sqrt{A_{\mathbf q}^2-B_{\mathbf q}^2},
\qquad
A_{\mathbf q}=4SJ_1-4SJ_2(1-\gamma_2)+D(2S-1),
\qquad
B_{\mathbf q}=4SJ_1\gamma_1,
\]

where \(\mathbf q\) is measured relative to the ordering vector,
\(A_{\mathbf q}\) and \(B_{\mathbf q}\) are the standard
linear-spin-wave coefficients, and \(S\) again denotes the spin quantum
number. As in the ferromagnetic case, the single-ion anisotropy enters
as the exact linear-spin-wave shift \(D(2S-1)\), which vanishes for
\(S=\tfrac12\). We use
\(\gamma_1=\tfrac12[\cos(2\pi q_h)+\cos(2\pi q_k)]\) and
\(\gamma_2=\cos(2\pi q_h)\cos(2\pi q_k)\) for \(q_h=q_k=H-0.5\).

At the software level, TAS-AI supports three backend families:

\begin{enumerate}
\def\labelenumi{\arabic{enumi}.}
\tightlist
\item
  \textbf{built-in analytical Python models} for rapid real-time
  operation,
\item
  \textbf{PySpinW} for SpinW-compatible calculations in more complex
  Hamiltonians,\textsuperscript{28} and
\item
  \textbf{Sunny.jl} as an optional advanced backend.\textsuperscript{29}
\end{enumerate}

Most results in this paper use the built-in analytical models. The main
exceptions are the analytic TAS-AI physics-only rows in the blind
benchmark table, which were run with a Sunny square-lattice backend, and
the PySpinW ground-truth benchmarks, which use PySpinW-generated data
plus corresponding TAS-AI runs.

\hypertarget{motion-aware-sequencing-and-mcts-batch-planning}{%
\subsubsection{3.7 Motion-aware sequencing and MCTS batch
planning}\label{motion-aware-sequencing-and-mcts-batch-planning}}

Motor motion enters the objective through the time denominator of the
acquisition score. For simplified \((H,E)\) trajectories, move time is
modeled as

\[
\mathrm{MoveTime}
=
\max\!\left(\frac{|\Delta H|}{v_H}, \frac{|\Delta E|}{v_E}\right)
+
t_{\mathrm{overhead}},
\]

where \(\Delta H\) and \(\Delta E\) are the proposed motor moves in
reciprocal-lattice and energy coordinates, \(v_H\) and \(v_E\) are the
corresponding motor speeds, and \(t_{\mathrm{overhead}}\) is a fixed
settling/readout overhead. This term converts information gain into
\textbf{information rate}, which is the quantity relevant to scarce beam
time.

The default queue builder is greedy and one-step. When measurements are
executed in short batches, however, trajectory ordering becomes path
dependent. TAS-AI therefore includes an optional \textbf{Monte Carlo
Tree Search} batch planner that evaluates short sequences and optimizes
information gained per total time.\textsuperscript{30,31}

\begin{center}\rule{0.5\linewidth}{0.5pt}\end{center}

\hypertarget{results}{%
\subsection{4. Results}\label{results}}

\hypertarget{blind-reconstruction-shows-why-hybrid-autonomy-is-necessary}{%
\subsubsection{4.1 Blind reconstruction shows why hybrid autonomy is
necessary}\label{blind-reconstruction-shows-why-hybrid-autonomy-is-necessary}}

We first evaluate the setting in which the controller is effectively
blind: it must discover the topology of the response surface before any
trustworthy model-based planning is possible. Figure 2 shows the
synthetic benchmark families and Figures 3--4 summarize the analytic and
PySpinW ground-truth benchmark results.

\begin{figure}
\centering
\includegraphics[width=6in,height=\textheight]{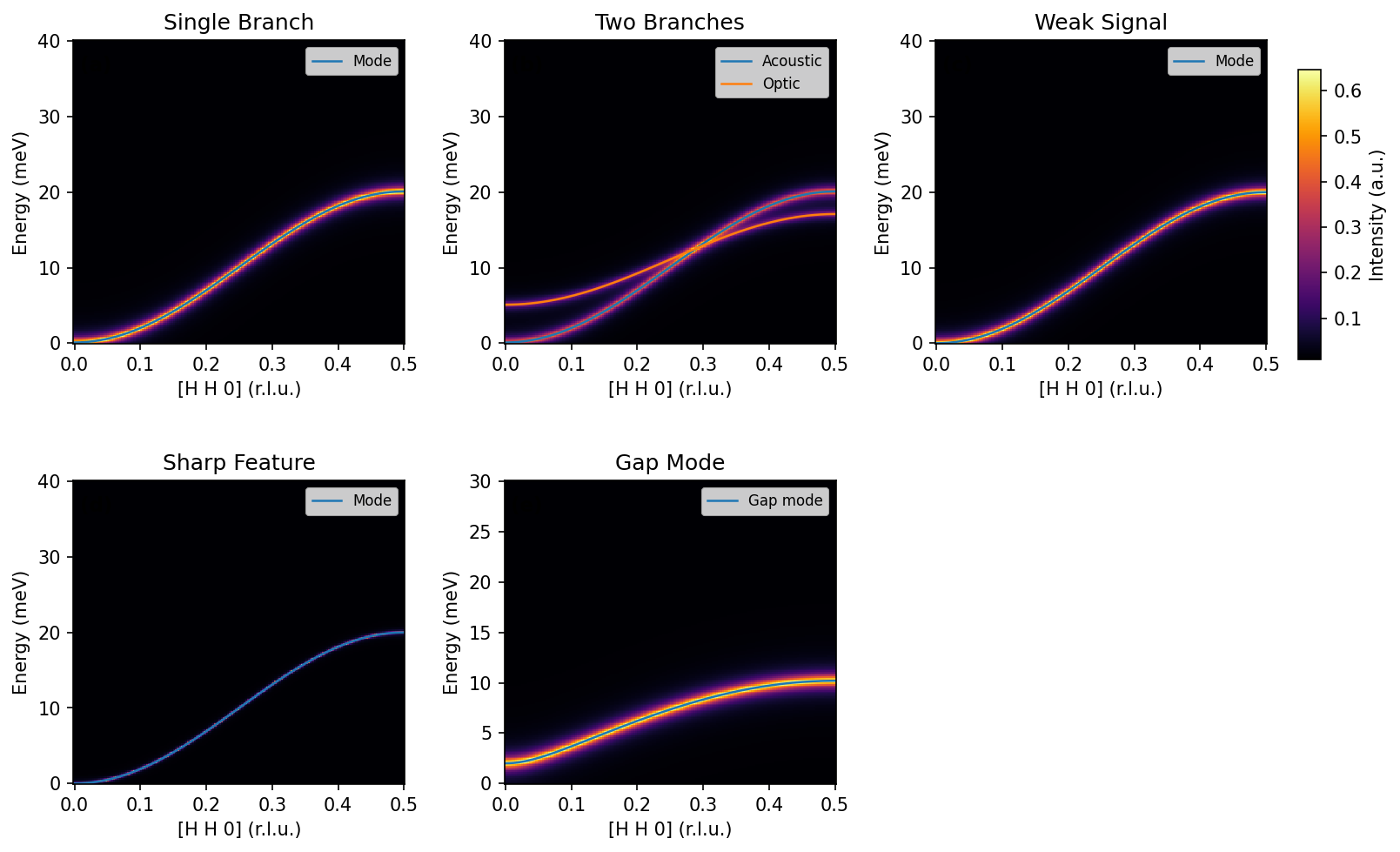}
\caption{Synthetic benchmark scenarios used to test discovery-oriented behavior: single branch, two branches, weak signal, sharp feature, and gap mode.}
\end{figure}

Table 1 reports the unified fair-benchmark summary. The reconstruction
metric is a truth-weighted mean absolute error evaluated on a fixed
reference grid in \((H,E)\) space, normalized so that a perfect
reconstruction scores zero and larger values indicate worse coverage
(the precise formula and implementation details are given in
Supplementary Note S2.1). A run is counted as successful when this error
falls below 0.20 within the measurement budget. The important point is
not which method is ``best'' in the abstract, but \textbf{which task is
being measured}. The global reconstruction metric primarily rewards
broad discovery coverage, so agnostic methods should be expected to
perform well. That is exactly what we observe in the analytic suite:
grid, random, and enhanced Log-GP usually cross the global error
threshold faster and more reliably than physics-only TAS-AI.

\begin{figure}
\centering
\includegraphics[width=6in,height=\textheight]{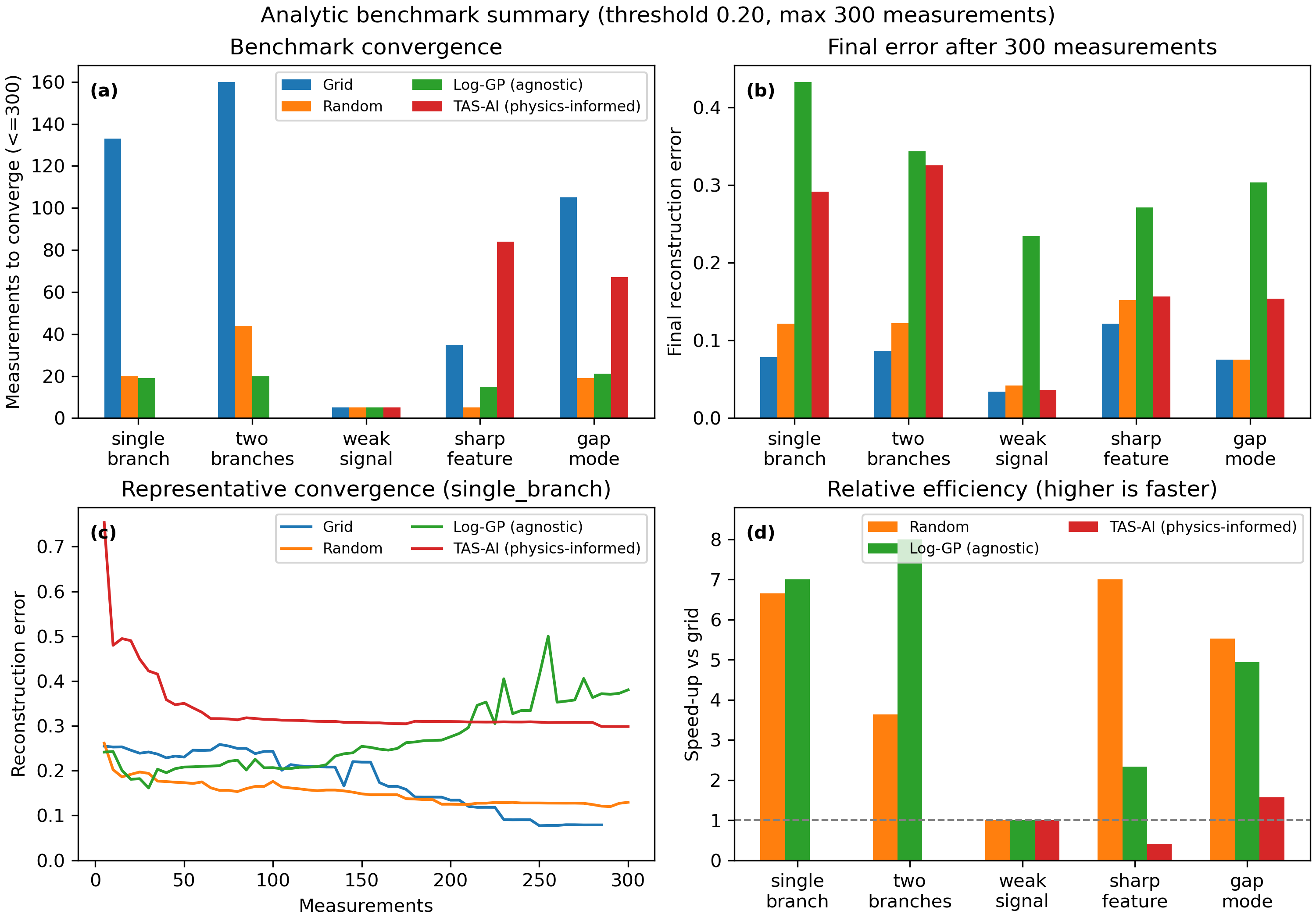}
\caption{Analytic blind-reconstruction benchmarks. Agnostic methods are favored by the global reconstruction metric because they are optimized for discovery rather than for parameter inference.}
\end{figure}

\begin{figure}
\centering
\includegraphics[width=6in,height=\textheight]{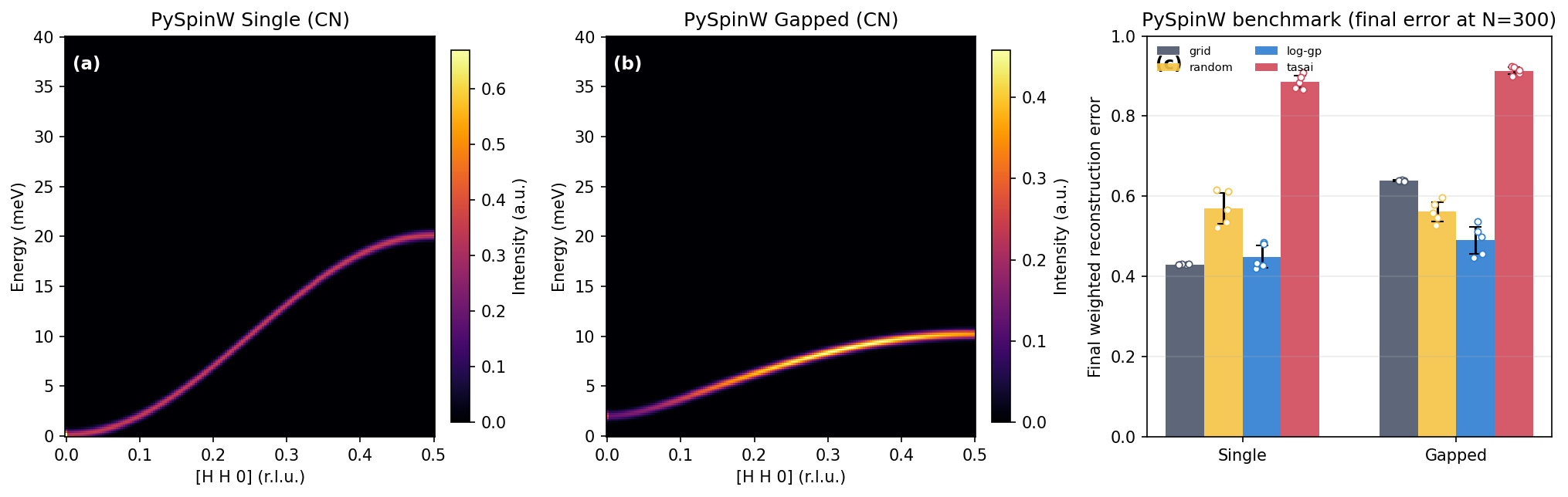}
\caption{PySpinW ground-truth benchmarks with Cooper--Nathans-derived energy broadening. Panels (a) and (b) show the two benchmark surfaces. Panel (c) reports mean final weighted reconstruction error after 300 measurements, with individual seed values overlaid. Under this stricter realism-heavy whole-window benchmark, none of the methods reaches the 0.20 threshold within budget; enhanced Log-GP is competitive with grid on the gapless case and performs best on the gapped case, while physics-only TAS-AI performs poorly on this blind global-mapping objective.}
\end{figure}

\emph{Table 1.} Fair benchmark summary with unified method settings.
Each cell reports success rate (X/5), median measurements-to-threshold
among successful runs {[}IQR{]}, and mean final error at budget across
all runs. Entries such as \texttt{5\ {[}5-5{]}} indicate that the
successful runs all reached threshold at the same measurement count, so
the IQR collapses to a single value. \texttt{N/A} indicates that no run
reached the reconstruction threshold within the fixed budget, so the
median measurements-to-threshold is undefined; in those cases, the
informative quantity is the final error at budget. Analytic rows use
threshold 0.20 and budget \(N=300\). For the corrected
PySpinW+Cooper-Nathans rows, no method reaches the 0.20 threshold within
budget. These results quantify \textbf{blind reconstruction}, not
downstream parameter inference.

\begin{longtable}[]{@{}lllll@{}}
\toprule
\begin{minipage}[b]{0.18\columnwidth}\raggedright
\textbf{Benchmark scenario}\strut
\end{minipage} & \begin{minipage}[b]{0.21\columnwidth}\raggedright
\textbf{Grid (succ; med {[}IQR{]}; err)}\strut
\end{minipage} & \begin{minipage}[b]{0.15\columnwidth}\raggedright
\textbf{Random}\strut
\end{minipage} & \begin{minipage}[b]{0.15\columnwidth}\raggedright
\textbf{Log-GP (enhanced)}\strut
\end{minipage} & \begin{minipage}[b]{0.17\columnwidth}\raggedright
\textbf{TAS-AI (physics-only)}\strut
\end{minipage}\tabularnewline
\midrule
\endhead
\begin{minipage}[t]{0.18\columnwidth}\raggedright
single\_branch\strut
\end{minipage} & \begin{minipage}[t]{0.21\columnwidth}\raggedright
5/5; 140 {[}140-140{]}; 0.079\strut
\end{minipage} & \begin{minipage}[t]{0.15\columnwidth}\raggedright
5/5; 15 {[}10-20{]}; 0.122\strut
\end{minipage} & \begin{minipage}[t]{0.15\columnwidth}\raggedright
5/5; 20 {[}20-20{]}; 0.433\strut
\end{minipage} & \begin{minipage}[t]{0.17\columnwidth}\raggedright
5/5; 85 {[}75-90{]}; 0.160\strut
\end{minipage}\tabularnewline
\begin{minipage}[t]{0.18\columnwidth}\raggedright
two\_branches\strut
\end{minipage} & \begin{minipage}[t]{0.21\columnwidth}\raggedright
5/5; 160 {[}160-160{]}; 0.087\strut
\end{minipage} & \begin{minipage}[t]{0.15\columnwidth}\raggedright
5/5; 35 {[}30-35{]}; 0.122\strut
\end{minipage} & \begin{minipage}[t]{0.15\columnwidth}\raggedright
5/5; 20 {[}20-20{]}; 0.343\strut
\end{minipage} & \begin{minipage}[t]{0.17\columnwidth}\raggedright
5/5; 150 {[}80-175{]}; 0.137\strut
\end{minipage}\tabularnewline
\begin{minipage}[t]{0.18\columnwidth}\raggedright
weak\_signal\strut
\end{minipage} & \begin{minipage}[t]{0.21\columnwidth}\raggedright
5/5; 5 {[}5-5{]}; 0.034\strut
\end{minipage} & \begin{minipage}[t]{0.15\columnwidth}\raggedright
5/5; 5 {[}5-5{]}; 0.042\strut
\end{minipage} & \begin{minipage}[t]{0.15\columnwidth}\raggedright
5/5; 5 {[}5-5{]}; 0.234\strut
\end{minipage} & \begin{minipage}[t]{0.17\columnwidth}\raggedright
5/5; 5 {[}5-5{]}; 0.074\strut
\end{minipage}\tabularnewline
\begin{minipage}[t]{0.18\columnwidth}\raggedright
sharp\_feature\strut
\end{minipage} & \begin{minipage}[t]{0.21\columnwidth}\raggedright
5/5; 35 {[}35-35{]}; 0.122\strut
\end{minipage} & \begin{minipage}[t]{0.15\columnwidth}\raggedright
5/5; 5 {[}5-5{]}; 0.152\strut
\end{minipage} & \begin{minipage}[t]{0.15\columnwidth}\raggedright
5/5; 15 {[}15-15{]}; 0.271\strut
\end{minipage} & \begin{minipage}[t]{0.17\columnwidth}\raggedright
5/5; 80 {[}68-150{]}; 0.173\strut
\end{minipage}\tabularnewline
\begin{minipage}[t]{0.18\columnwidth}\raggedright
gap\_mode\strut
\end{minipage} & \begin{minipage}[t]{0.21\columnwidth}\raggedright
5/5; 105 {[}105-105{]}; 0.075\strut
\end{minipage} & \begin{minipage}[t]{0.15\columnwidth}\raggedright
5/5; 15 {[}15-15{]}; 0.076\strut
\end{minipage} & \begin{minipage}[t]{0.15\columnwidth}\raggedright
4/5; 22 {[}10-25{]}; 0.303\strut
\end{minipage} & \begin{minipage}[t]{0.17\columnwidth}\raggedright
1/5; 70 {[}70-70{]}; 0.200\strut
\end{minipage}\tabularnewline
\begin{minipage}[t]{0.18\columnwidth}\raggedright
pyspinw\_single (CN, N=300)\strut
\end{minipage} & \begin{minipage}[t]{0.21\columnwidth}\raggedright
0/5; N/A; 0.430\strut
\end{minipage} & \begin{minipage}[t]{0.15\columnwidth}\raggedright
0/5; N/A; 0.571\strut
\end{minipage} & \begin{minipage}[t]{0.15\columnwidth}\raggedright
0/5; N/A; 0.449\strut
\end{minipage} & \begin{minipage}[t]{0.17\columnwidth}\raggedright
0/5; N/A; 0.885\strut
\end{minipage}\tabularnewline
\begin{minipage}[t]{0.18\columnwidth}\raggedright
pyspinw\_gapped (CN, N=300)\strut
\end{minipage} & \begin{minipage}[t]{0.21\columnwidth}\raggedright
0/5; N/A; 0.639\strut
\end{minipage} & \begin{minipage}[t]{0.15\columnwidth}\raggedright
0/5; N/A; 0.562\strut
\end{minipage} & \begin{minipage}[t]{0.15\columnwidth}\raggedright
0/5; N/A; 0.490\strut
\end{minipage} & \begin{minipage}[t]{0.17\columnwidth}\raggedright
0/5; N/A; 0.913\strut
\end{minipage}\tabularnewline
\bottomrule
\end{longtable}

This result is evidence that \textbf{discovery and inference are
different objectives}. A planner optimized for parameter information
rate is not the right tool for blind global mapping --- and that
separation is precisely the reason the hybrid workflow exists.

\hypertarget{physics-informed-planning-and-motion-aware-scheduling}{%
\subsubsection{4.2 Physics-informed planning and motion-aware
scheduling}\label{physics-informed-planning-and-motion-aware-scheduling}}

Figure 5 shows the controlled time-aware parameter-refinement study.
Here the Hamiltonian family is assumed known and the problem is no
longer blind discovery but parameter contraction under a realistic
wall-clock budget. The figure should therefore be read as a
refinement-stage demonstration. In this setting TAS-AI behaves as
intended: it reaches the target RMS threshold after 8 measurements and
about 170 s of elapsed experiment time, whereas the best competing
method in the representative run (random) reaches the same threshold
only after 17 measurements and about 542 s.

\begin{figure}
\centering
\includegraphics[width=6in,height=\textheight]{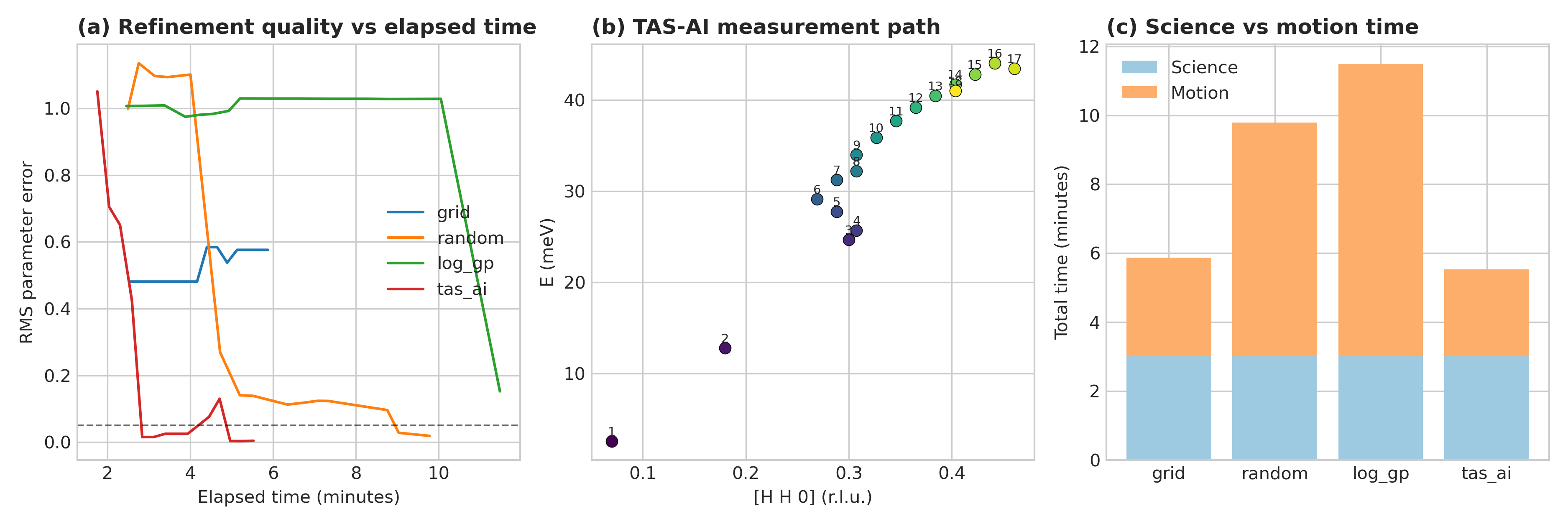}
\caption{Controlled time-aware parameter-refinement study for the square-lattice ferromagnet. Once the model family is fixed, the motion-aware physics-informed planner reaches the RMS threshold earliest by prioritizing parameter-sensitive measurements with high information rate. Panel (a) shows RMS error versus elapsed experiment time, panel (b) shows the TAS-AI path through \((H,E)\) space, and panel (c) separates science counting from motion overhead. In the representative run shown here, TAS-AI reaches threshold in 170 s while the random baseline does so only after 542 s; grid and Log-GP do not converge within the same budget.}
\end{figure}

Figure 6 then shows the in-loop NN-vs-\(J_1\)-\(J_2\) discrimination
test. The controller reaches decisive AIC-derived evidence within the
first few batch updates by selecting points near the band edge and the
intermediate-\(H\) shoulder where the models separate most strongly.

\begin{figure}
\centering
\includegraphics[width=6in,height=\textheight]{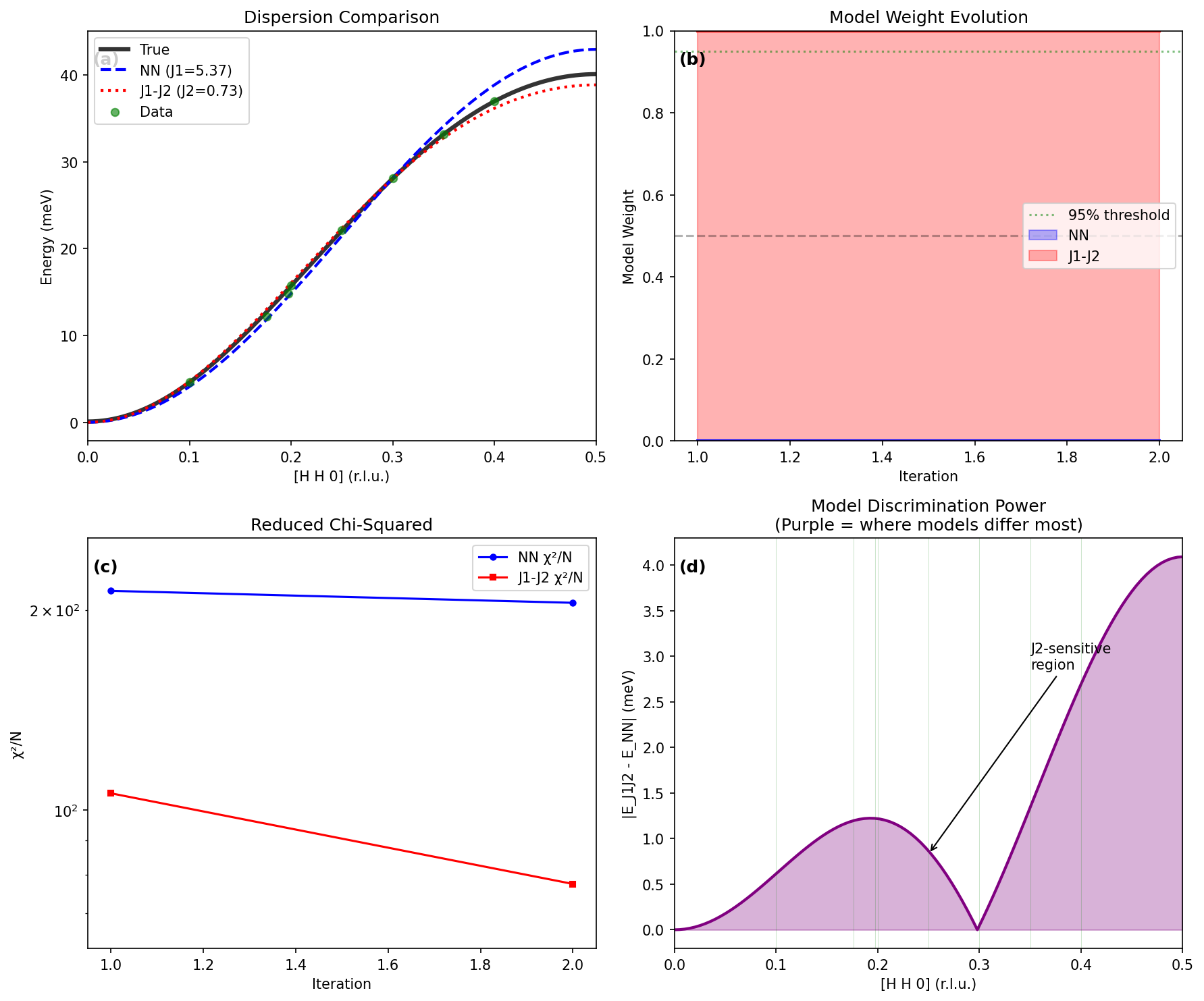}
\caption{In-loop Hamiltonian discrimination. The x-axis counts cumulative measurements, including the seeded points used to initialize the comparison; thus the decisive jump occurs within eight total measurements rather than eight post-seed updates. The AIC-derived evidence ratio becomes decisive in the representative run because the planner targets the regions where the competing dispersions diverge most strongly. In panel (d), the green vertical guides mark the measurement locations used for the corresponding discrimination trace.}
\end{figure}

The speed of this discrimination depends on the chemically informed
prior (§3.3); with flat priors the correct model is still identified,
but more measurements are needed to reach the decisive threshold
(Supplementary Note S5.4).

This result illustrates the central design principle: once discovery has
done its job, model-aware planning is no longer competing on the same
axis as a global mapper. It is competing on the axis that matters for
physical interpretation --- how fast the experiment can determine the
correct model and tighten the relevant posteriors.

An autonomous loop that optimizes information gain without accounting
for instrument motion can still waste beam time. Figure 7 and Table 2
isolate this effect in a \textbf{controlled scheduling study}: all
methods are given the same fixed set of scientifically relevant
candidate measurements, and only the ordering policy varies.
Motion-aware ordering reduces total run time from 88 minutes to 60
minutes for the same 50 minutes of science counting.

\begin{figure}
\centering
\includegraphics[width=6in,height=\textheight]{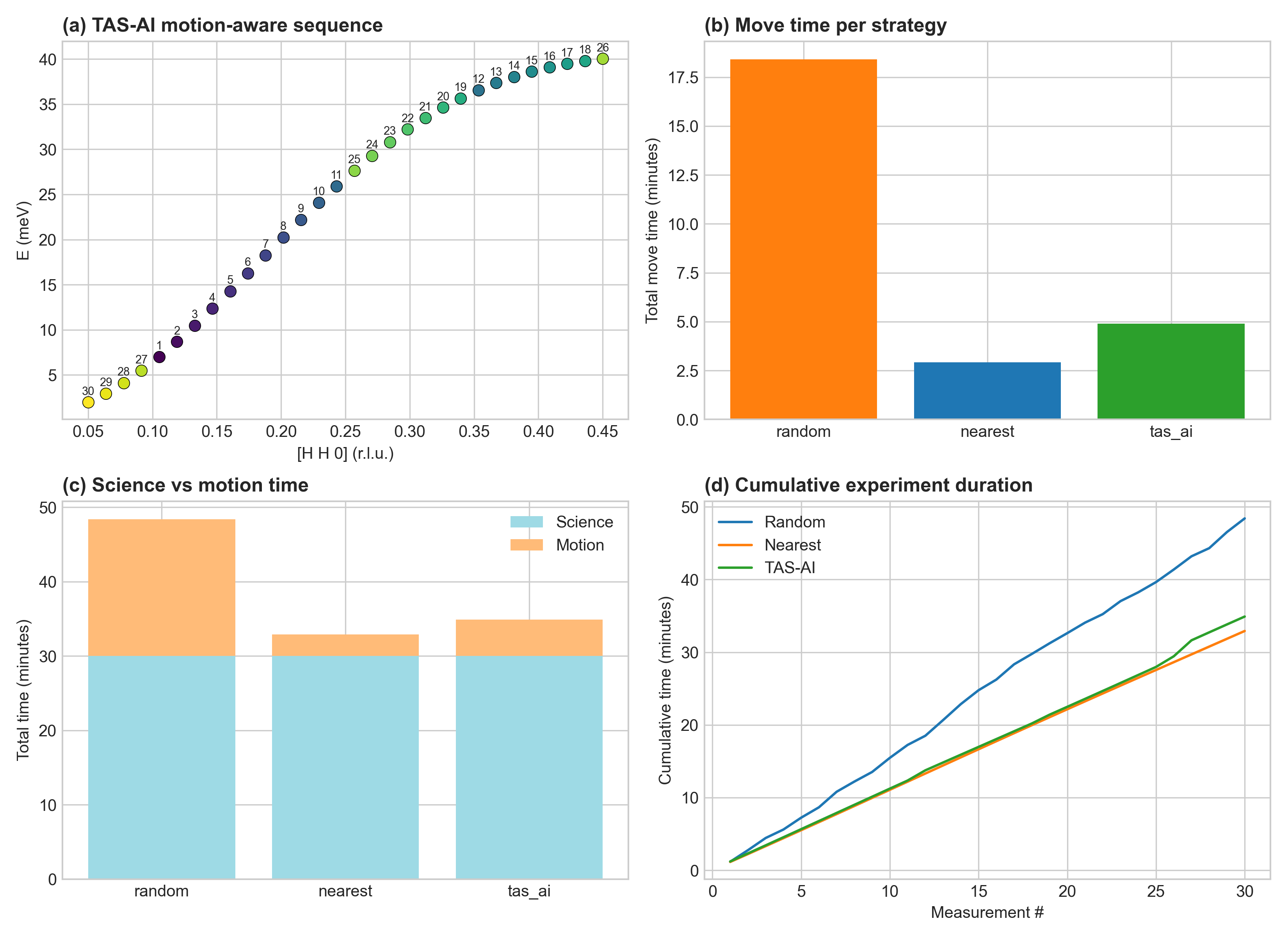}
\caption{Controlled motion-aware scheduling diagnostics on a fixed candidate set. Panel (a) shows the fixed candidate measurements colored by their position in the executed schedule, so the color scale encodes traversal order rather than intensity. Panels (b)-(d) then separate the corresponding cumulative path length, elapsed time, and per-point timing summaries. The figure is intended as a scheduling diagnostic, not as a second benchmark of adaptive discovery.}
\end{figure}

\emph{Table 2.} Motion-aware scheduling timings for the controlled
fixed-candidate scenario in Figure 7. This table isolates route-ordering
effects rather than benchmarking full adaptive discovery policies.

\begin{longtable}[]{@{}lllll@{}}
\toprule
\textbf{Strategy} & \textbf{Science Time} & \textbf{Move Time} &
\textbf{Total} & \textbf{Efficiency}\tabularnewline
\midrule
\endhead
Random order & 50 min & 38 min & 88 min & 57\%\tabularnewline
Nearest neighbor & 50 min & 22 min & 72 min & 69\%\tabularnewline
TAS-AI optimized & 50 min & 10 min & 60 min & 83\%\tabularnewline
\bottomrule
\end{longtable}

This effect is not a secondary implementation detail. On a TAS
instrument, a 30--60 s move can be comparable to a 60 s count, so
wall-clock throughput depends on both the physics utility and the route
taken through phase space.

Batched trajectories introduce an additional path dependence. Figure 8
shows that MCTS outperforms one-step greedy ordering in motion-dominated
regimes by explicitly evaluating short candidate sequences. A systematic
multi-seed evaluation of the MCTS benefit across broader motion models
and batch horizons is beyond the scope of this paper but is a natural
follow-up.

\begin{figure}
\centering
\includegraphics[width=6in,height=\textheight]{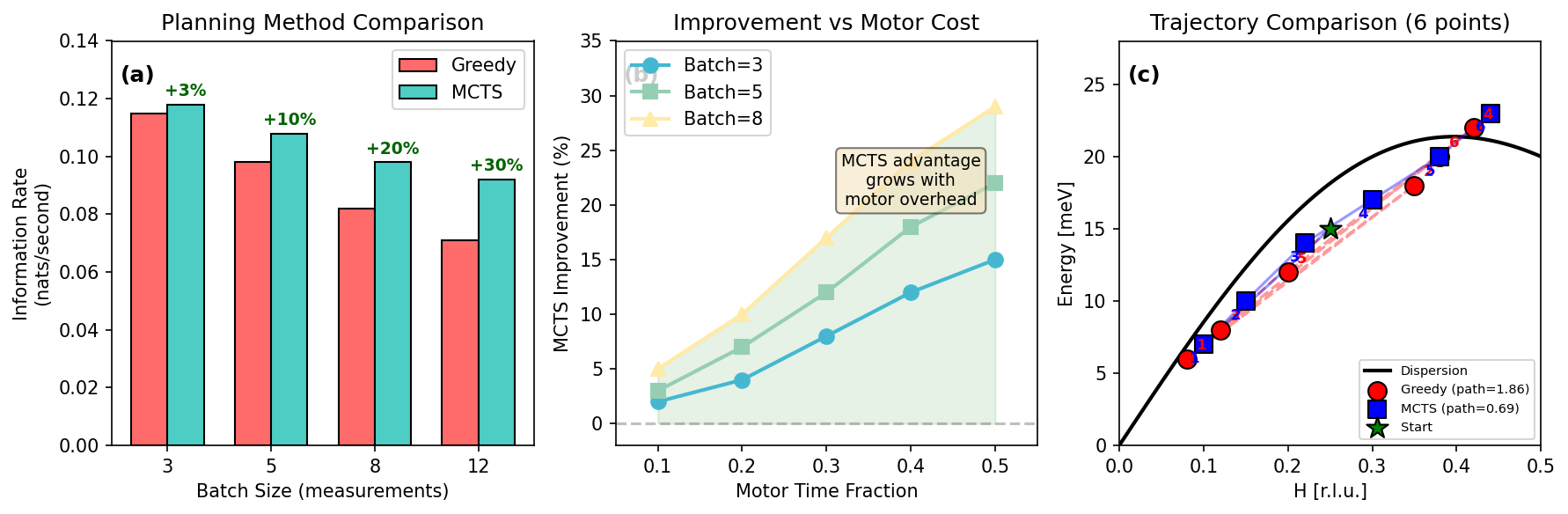}
\caption{MCTS batch planning reduces path inefficiency relative to one-step greedy ordering when motion dominates the cost budget.}
\end{figure}

\hypertarget{the-full-hybrid-handoff-in-an-integrated-run}{%
\subsubsection{4.3 The full hybrid handoff in an integrated
run}\label{the-full-hybrid-handoff-in-an-integrated-run}}

The core design principle of the paper is best seen in a single
integrated run. Figure 9 shows the handoff from enhanced Log-GP
discovery to physics-informed inference, with checkpoints saved at each
controller-phase boundary. The archived run uses a fixed three-phase
schedule for exposition --- 13 coarse grid points, 15 enhanced Log-GP
active points, then physics refinement from measurement 29 --- so that
the control transition is visible at a glance; the automatic handoff
trigger of §3.2 selects a similar transition point from the same seeded
survey. The posterior remains non-decisive through the agnostic stage
and sharpens to decisive support for the full model only once physics
refinement begins. The posterior evolution is phase-labeled so the
reader can see that the control logic is not monolithic: the system
changes acquisition strategy as the task changes.

\begin{figure}
\centering
\includegraphics[width=6in,height=\textheight]{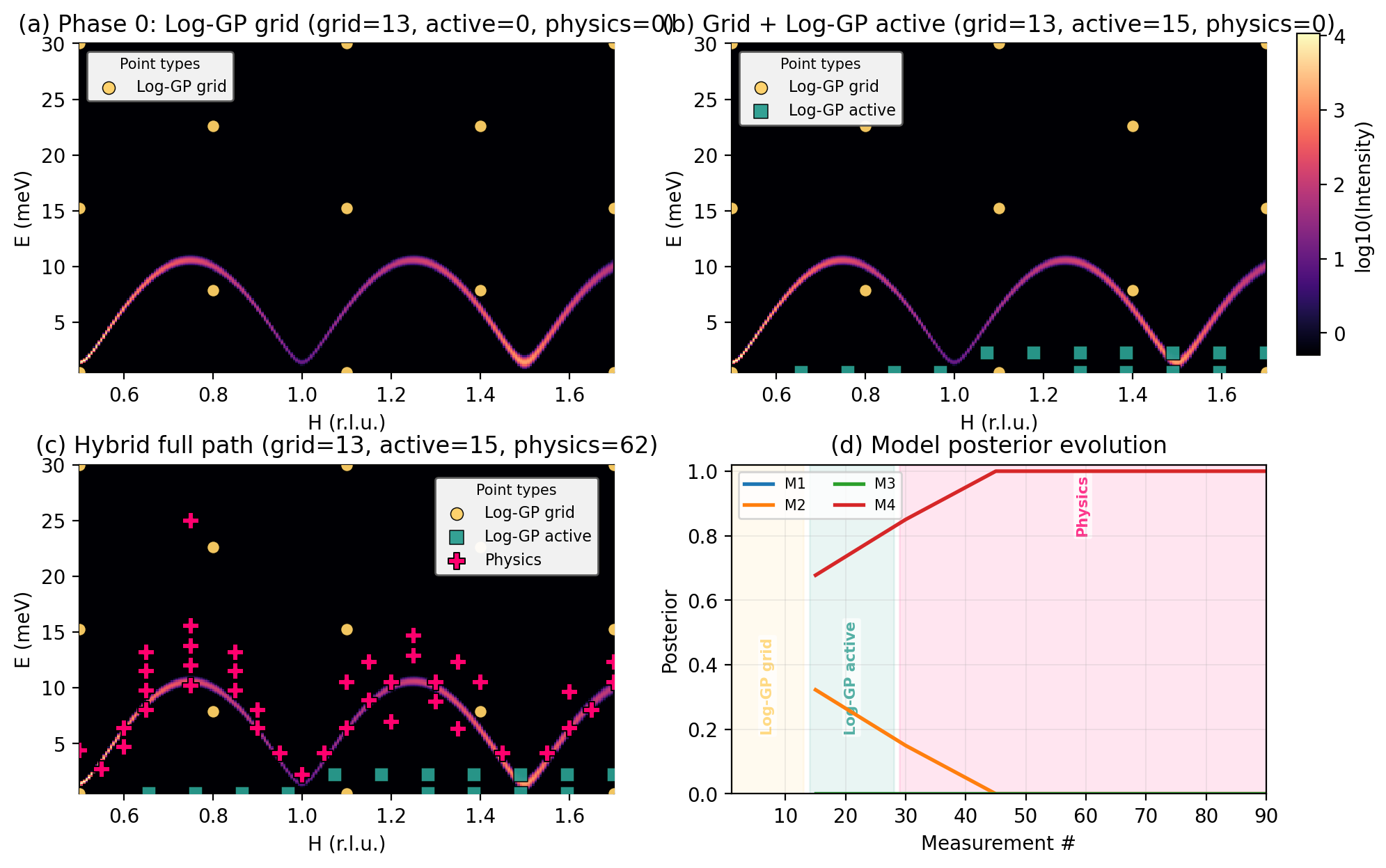}
\caption{Hybrid handoff from agnostic discovery to physics-informed inference. The agnostic front end uses the enhanced Log-GP policy (§3.1). In panel (d), the colored vertical bands mark the contiguous controller phases: coarse grid discovery, active Log-GP remapping, and physics-informed refinement. The figure makes the control transition explicit rather than mixing all points into a single trajectory.}
\end{figure}

This integrated run is the clearest evidence for the manuscript's main
thesis: the agnostic front end is not an auxiliary convenience, and the
physics-informed back end is not a drop-in replacement for discovery.
The power comes from \textbf{deploying each controller in the regime
where its inductive bias matches the task}.

\hypertarget{computational-performance}{%
\subsubsection{4.4 Computational
performance}\label{computational-performance}}

The core algorithmic components of TAS-AI operate comfortably within
real-time beamline latencies. Typical per-iteration timings are
approximately 0.5 ms for dispersion evaluation over 100 points, 2 ms for
intensity calculation, 10 ms for acquisition optimization, 100 ms for
fast local fitting, and 500 ms for a heavier 1000-sample MCMC stage when
triggered.

The distinction between algorithmic and benchmark-harness timing
matters. The \textbf{sub-second} timing refers to the inference and
planning latency of the numerical controller itself. Supplementary Table
S1 instead reports planner-side \texttt{mean\_time\_per\_suggestion} for
the benchmark harness: approximately \texttt{0.85-\/-0.97\ s} for
enhanced Log-GP and \texttt{0.016\ s} for TAS-AI (physics). Both are
relevant: the former for real beamline feasibility, the latter for
reproducibility and benchmark cost accounting.

We now turn to a qualitatively different challenge: what happens when
the posterior-weighted planner itself becomes the bottleneck to correct
model selection.

\begin{center}\rule{0.5\linewidth}{0.5pt}\end{center}

\hypertarget{mitigating-algorithmic-myopia-via-strategic-audit}{%
\subsection{5. Mitigating algorithmic myopia via strategic
audit}\label{mitigating-algorithmic-myopia-via-strategic-audit}}

Having identified silent-data posterior lock-in in §3.4, we now present
a constrained audit layer designed to mitigate it. The section describes
the implementation (§5.1), reports a pilot closed-loop demonstration
(§5.2), and presents two controlled ablation benchmarks that isolate the
lock-in failure mode (§5.3).

\hypertarget{constrained-llm-oversight}{%
\subsubsection{5.1 Constrained LLM
oversight}\label{constrained-llm-oversight}}

In the present implementation, the audit role is instantiated by an
\textbf{LLM committee} operating under strict constraints. The committee
does not see the full hidden numerical state of the planner. Instead it
receives a compact prompt packet containing only the information needed
for strategic auditing: a short recent-history table of measured
\((H,E,I)\) points; the current measurement count and batch index; the
allowed mode choices; a narrow description of the current ambiguity
(generated automatically from loop state, not typed by a human during
the run); and the hard execution constraints. Those constraints include
the allowed reciprocal-space window, the allowed energy window, the
requirement that responses be returned as strict JSON, and the rule that
the committee may not alter likelihoods, fit models, or override
kinematic vetoes.

The committee is then allowed to do only two things:

\begin{enumerate}
\def\labelenumi{\arabic{enumi}.}
\tightlist
\item
  choose the next \textbf{mode} (\texttt{loggp\_active} or
  \texttt{physics}), and/or\\
\item
  nominate a \textbf{small number of tactical audit points} from the
  bounded candidate menu already constructed by the numerical planner.
\end{enumerate}

Operationally, two proposer models generate independent candidate JSON
payloads from the same redacted prompt, and a third decider model
selects between them or falls back to the safest valid option. The LLM
therefore does not emit arbitrary continuous instrument coordinates; it
chooses among bounded routing options and menu-listed audit probes that
have already passed the numerical guardrails. In the current watcher
implementation, the provider pool is Claude Code (default model: Opus
4.5), Gemini CLI (default model: Gemini 3), and Codex CLI pinned to
\texttt{gpt-5.2-codex}; the decider rotates across those providers by
batch unless explicitly pinned, to reduce stylistic bias and provide
failure containment. Additional reproducibility details are given in
Supplementary Note S5.

Everything else remains symbolic and numerical. The LLM does
\textbf{not} fit Hamiltonians, update model weights, alter the
likelihood, bypass kinematic checks, or reorder the entire queue. The
Bayesian engine still computes the posteriors, and the instrument
planner still enforces all accessibility and safety constraints. The LLM
therefore functions as a \textbf{strategic auditor}, not as a
replacement for the inference loop.

This division of labor keeps the system scientifically interpretable
while allowing a flexible source of tactical hypotheses when the
internal utility becomes too self-confirming. In effect, the audit layer
can say: \emph{before spending another batch exploiting the current
leader, allocate one or two measurements to the falsification probes
that the greedy utility is currently underweighting.}

\hypertarget{pilot-closed-loop-demonstration}{%
\subsubsection{5.2 Pilot closed-loop
demonstration}\label{pilot-closed-loop-demonstration}}

Figure 10 shows a full 90-measurement closed-loop run with the LLM audit
layer active. The test system is a square-lattice AFM
\(J_1\)-\(J_2\)-\(D\) model centered on
\(\mathbf Q_{\mathrm{AF}}=(0.5,0.5,0)\) with four nested candidates:
NN-only (\(M_1\)), NN+\(D\) (\(M_2\)), NN+\(J_2\) (\(M_3\)), and the
full \(J_1\)+\(J_2\)+\(D\) model (\(M_4\)). The synthetic data are
generated from \(M_4\). Starting from the enhanced Log-GP coarse grid
and active warm start, the overseer alternates between agnostic
remapping and physics batches while remaining inside the same
constraints as the non-LLM loop. The final fit selects the full model
decisively (Table 3).

\begin{figure}
\centering
\includegraphics[width=6in,height=\textheight]{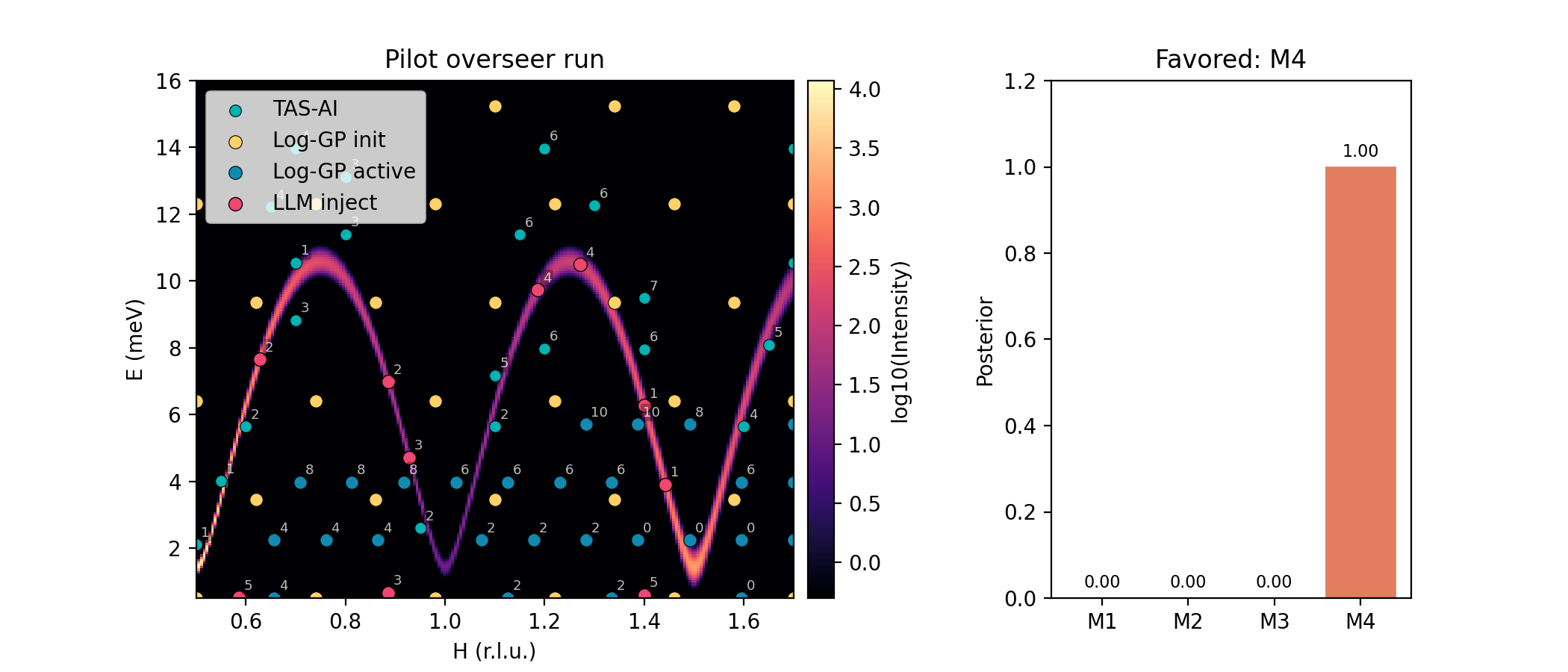}
\caption{Pilot LLM-audited closed-loop run (90 measurements). The left panel shows the executed measurements over the synthetic intensity map, with enhanced Log-GP grid, enhanced Log-GP active, and physics points shown separately. The right panel shows the final model posteriors. The overseer operates only at the routing layer while preserving the same numerical posteriors, kinematic constraints, and safety checks as the non-LLM loop.}
\end{figure}

\emph{Table 3.} Closed-loop pilot fit summary. Uncertainties are Laplace
approximations from the local Hessian; empirical coverage is
substantially below nominal (Supplementary Note S3.4), so these should
be read as local error surrogates rather than calibrated credible
intervals.

\begin{longtable}[]{@{}lrrllll@{}}
\toprule
\begin{minipage}[b]{0.18\columnwidth}\raggedright
\textbf{Scenario}\strut
\end{minipage} & \begin{minipage}[b]{0.04\columnwidth}\raggedleft
\textbf{N}\strut
\end{minipage} & \begin{minipage}[b]{0.12\columnwidth}\raggedleft
\textbf{M4 posterior}\strut
\end{minipage} & \begin{minipage}[b]{0.11\columnwidth}\raggedright
\textbf{J₁ (meV)}\strut
\end{minipage} & \begin{minipage}[b]{0.11\columnwidth}\raggedright
\textbf{J₂ (meV)}\strut
\end{minipage} & \begin{minipage}[b]{0.13\columnwidth}\raggedright
\textbf{D (meV)}\strut
\end{minipage} & \begin{minipage}[b]{0.11\columnwidth}\raggedright
\textbf{Bkg (arb.)}\strut
\end{minipage}\tabularnewline
\midrule
\endhead
\begin{minipage}[t]{0.18\columnwidth}\raggedright
True\strut
\end{minipage} & \begin{minipage}[t]{0.04\columnwidth}\raggedleft
---\strut
\end{minipage} & \begin{minipage}[t]{0.12\columnwidth}\raggedleft
---\strut
\end{minipage} & \begin{minipage}[t]{0.11\columnwidth}\raggedright
1.25\strut
\end{minipage} & \begin{minipage}[t]{0.11\columnwidth}\raggedright
0.20\strut
\end{minipage} & \begin{minipage}[t]{0.13\columnwidth}\raggedright
0.020\strut
\end{minipage} & \begin{minipage}[t]{0.11\columnwidth}\raggedright
0.50\strut
\end{minipage}\tabularnewline
\begin{minipage}[t]{0.18\columnwidth}\raggedright
Non-LLM (no symmetry)\strut
\end{minipage} & \begin{minipage}[t]{0.04\columnwidth}\raggedleft
87\strut
\end{minipage} & \begin{minipage}[t]{0.12\columnwidth}\raggedleft
\textgreater0.999\strut
\end{minipage} & \begin{minipage}[t]{0.11\columnwidth}\raggedright
1.247 ± 0.004\strut
\end{minipage} & \begin{minipage}[t]{0.11\columnwidth}\raggedright
0.198 ± 0.004\strut
\end{minipage} & \begin{minipage}[t]{0.13\columnwidth}\raggedright
0.0201 ± 0.0001\strut
\end{minipage} & \begin{minipage}[t]{0.11\columnwidth}\raggedright
0.476 ± 0.011\strut
\end{minipage}\tabularnewline
\begin{minipage}[t]{0.18\columnwidth}\raggedright
Symmetry-seeded (no LLM)\strut
\end{minipage} & \begin{minipage}[t]{0.04\columnwidth}\raggedleft
86\strut
\end{minipage} & \begin{minipage}[t]{0.12\columnwidth}\raggedleft
\textgreater0.999\strut
\end{minipage} & \begin{minipage}[t]{0.11\columnwidth}\raggedright
1.250 ± 0.004\strut
\end{minipage} & \begin{minipage}[t]{0.11\columnwidth}\raggedright
0.199 ± 0.004\strut
\end{minipage} & \begin{minipage}[t]{0.13\columnwidth}\raggedright
0.0199 ± 0.0001\strut
\end{minipage} & \begin{minipage}[t]{0.11\columnwidth}\raggedright
0.485 ± 0.010\strut
\end{minipage}\tabularnewline
\begin{minipage}[t]{0.18\columnwidth}\raggedright
LLM-audited (no symmetry)\strut
\end{minipage} & \begin{minipage}[t]{0.04\columnwidth}\raggedleft
90\strut
\end{minipage} & \begin{minipage}[t]{0.12\columnwidth}\raggedleft
\textgreater0.999\strut
\end{minipage} & \begin{minipage}[t]{0.11\columnwidth}\raggedright
1.256 ± 0.010\strut
\end{minipage} & \begin{minipage}[t]{0.11\columnwidth}\raggedright
0.206 ± 0.009\strut
\end{minipage} & \begin{minipage}[t]{0.13\columnwidth}\raggedright
0.0211 ± 0.0007\strut
\end{minipage} & \begin{minipage}[t]{0.11\columnwidth}\raggedright
0.485 ± 0.010\strut
\end{minipage}\tabularnewline
\bottomrule
\end{longtable}

The purpose of this pilot is to demonstrate that the audit layer
integrates into the full workflow without degrading model selection or
parameter recovery. All three runs reach decisive support for \(M_4\).
The LLM-audited run shows modestly wider parameter uncertainties (e.g.,
\(\pm 0.010\) vs.~\(\pm 0.004\) on \(J_1\)) because the overseer diverts
some measurements toward falsification probes rather than parameter
refinement --- a deliberate trade-off between refinement precision and
model-selection robustness that the ablations in §5.3 examine in more
detail.

\hypertarget{escaping-posterior-lock-in-targeted-ablations}{%
\subsubsection{5.3 Escaping posterior lock-in: targeted
ablations}\label{escaping-posterior-lock-in-targeted-ablations}}

The pilot demonstration shows that the audit layer integrates into the
full workflow. We now isolate the lock-in failure mode directly through
two controlled ablation benchmarks of increasing physical complexity.

\hypertarget{ghost-optic-ablation}{%
\paragraph{5.3.1 Ghost-optic ablation}\label{ghost-optic-ablation}}

The first ablation uses a deliberately minimal fixed-\(Q\) analytic
benchmark in which the current Bayesian leader is a one-branch model
while the true spectrum contains an additional weak secondary branch
carrying only a small fraction of the dominant spectral weight.
Concretely, the ghost-optic benchmark is a fixed-\(Q\) two-Lorentzian
toy spectrum over \(E\in[0,20]\) with a dominant acoustic peak at
\(E=5\), a weak optic peak at \(E=15\) carrying 5\% of the acoustic
amplitude, shared linewidth \(\gamma=0.5\), and additive background 0.1;
the comparator omits the optic term. The common seed samples only the
bright acoustic peak at \(E=\{4.25,4.75,5.25,5.75\}\), leaving the optic
region initially untested.

\begin{figure}
\centering
\includegraphics[width=6in,height=\textheight]{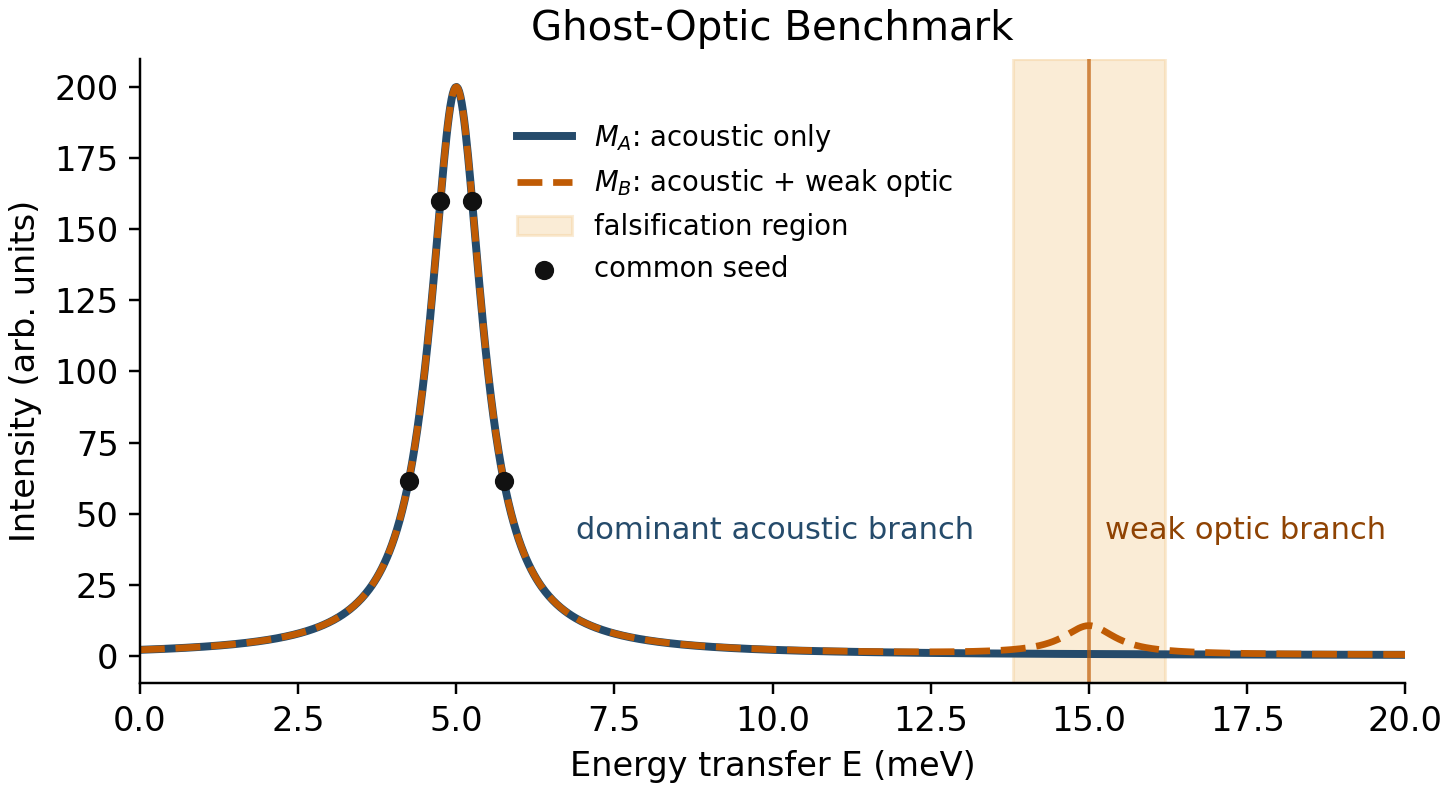}
\caption{Ghost-optic benchmark schematic used in the Section 5.3.1 audit ablation. The acoustic-only comparator (\(M_A\)) and the acoustic+optic truth (\(M_B\)) agree on the dominant bright branch near \(E=5\) but differ through a weak secondary optic feature near \(E=15\). Black markers show the common seed measurements, which cover only the acoustic branch and therefore leave the falsification region initially untested.}
\end{figure}

From that common seed, we compare four policies:

\begin{itemize}
\tightlist
\item
  \texttt{None}: fixed seed followed by pure refinement of the current
  leader.
\item
  \texttt{Log-GP}: fixed seed followed by a 1D GP variance explorer,
  that is, the bare Log-GP variant (§3.1), without the taper/clamp
  safeguards of the enhanced policy.
\item
  \texttt{Max-disagreement}: a deterministic top-two falsification rule
  that selects the kinematically accessible candidate with the largest
  intensity disagreement between the current leading and runner-up
  models, subject to the same bounded menu and injection budget as all
  other policies.
\item
  \texttt{LLM}: fixed seed followed by the same refinement loop plus the
  LLM audit layer, using the same shared candidate menu and strict JSON
  contract as the main overseer.
\end{itemize}

\emph{Table 4.} Ghost-optic audit ablation. The benchmark uses a fixed
sparse seed around the bright branch and compares a pure refinement
baseline (None), a bare Log-GP explorer (Log-GP), a deterministic
max-disagreement falsification rule (Max-disagreement), and the LLM
audit policy (LLM). Unlike Figures 9 and 10, this benchmark does not use
the full enhanced Log-GP discovery policy.

\begin{longtable}[]{@{}lrrrr@{}}
\toprule
\textbf{Policy} & \textbf{Time to decisive} & \textbf{Wrong-leader
dwell} & \textbf{Falsif. fraction} & \textbf{Success}\tabularnewline
\midrule
\endhead
None & 34 & 25 & 0 & 1\tabularnewline
Log-GP & 29 & 20 & 0.75 & 1\tabularnewline
Max-disagreement & 6 & 0 & 1 & 1\tabularnewline
LLM & 9 & 0 & 1 & 1\tabularnewline
\bottomrule
\end{longtable}

All four one-seed policies eventually recover in this minimal analytic
setting, but they do so on very different timescales. The
refinement-only baseline (None) spends 25 measurements remeasuring the
bright acoustic feature before reaching decisive correct selection; the
bare Log-GP explorer (Log-GP) reduces that dwell by allocating more
falsification-oriented batches; and both Max-disagreement and the LLM
audit eliminate wrong-leader dwell entirely. The deterministic
max-disagreement rule is faster than the LLM in this benchmark (decisive
at 6 rather than 9 measurements), reinforcing the conclusion that the
gain comes from the \textbf{falsification principle} rather than from
the specific implementation. The benchmark isolates the failure mode
cleanly: posterior lock-in on a bright feature delays falsification even
when the missing signal is physically simple and kinematically
accessible.

\hypertarget{bilayer-ferromagnet-ablation}{%
\paragraph{5.3.2 Bilayer ferromagnet
ablation}\label{bilayer-ferromagnet-ablation}}

To move beyond the minimal ghost-optic setting, we implemented a
square-lattice bilayer ferromagnet backend in which the acoustic branch
remains bright while a weak \(L\)-suppressed optic branch provides the
falsifying signal. The single-branch comparator shares the same
\(L\)-dependent acoustic weight as the bilayer truth, so the two models
differ only through the presence or absence of the optic branch. We
report a separate \texttt{optic\_region\_hit\_fraction} metric that
counts batches containing at least one measurement within a narrow
tolerance of the optic branch.

This benchmark uses the same shared action space as the full hybrid
loop: all controllers choose between bare Log-GP remapping
(\texttt{loggp\_active}) and \texttt{physics} refinement, and differ
only in whether that choice is made by a deterministic rule or by the
LLM overseer. As in Table 4, \texttt{loggp\_active} refers to the bare
Log-GP variant, not the full enhanced Log-GP policy of Figures 9 and 10.

\begin{itemize}
\tightlist
\item
  \texttt{None}: fixed seed followed by physics refinement only.
\item
  \texttt{Hybrid}: fixed seed followed by the shared action space
  (switch between \texttt{loggp\_active} and \texttt{physics}), with
  mode chosen by a deterministic non-LLM rule and no tactical audit
  injections.
\item
  \texttt{Max-disagreement}: same shared action space as Hybrid, but
  audit injections are selected by the deterministic top-two
  max-disagreement rule described in §5.3.1.
\item
  \texttt{LLM}: fixed seed followed by the same action space, but with
  the LLM overseer choosing the mode and optionally adding up to two
  menu-selected falsification probes.
\end{itemize}

In the reported one-seed comparison, the deterministic Hybrid router
uses the same menu of allowed actions and has no access to the true
model. It uses simple threshold rules on posterior entropy,
falsification-region coverage, and posterior margin to decide when to
switch modes, with a forced periodic \texttt{loggp\_active} batch to
guarantee minimum exploration. The precise trigger values are given in
Supplementary Note S5.

\emph{Table 5.} Bilayer ferromagnet audit ablation using the shared
action space. The falsification-probe fraction counts batches that
include at least one true falsification-region measurement, whether
reached incidentally by exploration or deliberately by audit injection.
The optic-region-hit fraction counts batches that actually sample the
weak optic branch region.

\begin{longtable}[]{@{}lrrrrr@{}}
\toprule
\begin{minipage}[b]{0.08\columnwidth}\raggedright
\textbf{Policy}\strut
\end{minipage} & \begin{minipage}[b]{0.16\columnwidth}\raggedleft
\textbf{Time to decisive}\strut
\end{minipage} & \begin{minipage}[b]{0.17\columnwidth}\raggedleft
\textbf{Wrong-leader dwell}\strut
\end{minipage} & \begin{minipage}[b]{0.16\columnwidth}\raggedleft
\textbf{Falsif. fraction}\strut
\end{minipage} & \begin{minipage}[b]{0.17\columnwidth}\raggedleft
\textbf{Optic-hit fraction}\strut
\end{minipage} & \begin{minipage}[b]{0.09\columnwidth}\raggedleft
\textbf{Success}\strut
\end{minipage}\tabularnewline
\midrule
\endhead
\begin{minipage}[t]{0.08\columnwidth}\raggedright
None\strut
\end{minipage} & \begin{minipage}[t]{0.16\columnwidth}\raggedleft
23\strut
\end{minipage} & \begin{minipage}[t]{0.17\columnwidth}\raggedleft
5\strut
\end{minipage} & \begin{minipage}[t]{0.16\columnwidth}\raggedleft
0\strut
\end{minipage} & \begin{minipage}[t]{0.17\columnwidth}\raggedleft
0.56\strut
\end{minipage} & \begin{minipage}[t]{0.09\columnwidth}\raggedleft
1\strut
\end{minipage}\tabularnewline
\begin{minipage}[t]{0.08\columnwidth}\raggedright
Hybrid\strut
\end{minipage} & \begin{minipage}[t]{0.16\columnwidth}\raggedleft
8\strut
\end{minipage} & \begin{minipage}[t]{0.17\columnwidth}\raggedleft
0\strut
\end{minipage} & \begin{minipage}[t]{0.16\columnwidth}\raggedleft
0.17\strut
\end{minipage} & \begin{minipage}[t]{0.17\columnwidth}\raggedleft
0.50\strut
\end{minipage} & \begin{minipage}[t]{0.09\columnwidth}\raggedleft
1\strut
\end{minipage}\tabularnewline
\begin{minipage}[t]{0.08\columnwidth}\raggedright
Max-disagreement\strut
\end{minipage} & \begin{minipage}[t]{0.16\columnwidth}\raggedleft
5\strut
\end{minipage} & \begin{minipage}[t]{0.17\columnwidth}\raggedleft
0\strut
\end{minipage} & \begin{minipage}[t]{0.16\columnwidth}\raggedleft
0.64\strut
\end{minipage} & \begin{minipage}[t]{0.17\columnwidth}\raggedleft
0.64\strut
\end{minipage} & \begin{minipage}[t]{0.09\columnwidth}\raggedleft
1\strut
\end{minipage}\tabularnewline
\begin{minipage}[t]{0.08\columnwidth}\raggedright
LLM\strut
\end{minipage} & \begin{minipage}[t]{0.16\columnwidth}\raggedleft
8\strut
\end{minipage} & \begin{minipage}[t]{0.17\columnwidth}\raggedleft
0\strut
\end{minipage} & \begin{minipage}[t]{0.16\columnwidth}\raggedleft
0.50\strut
\end{minipage} & \begin{minipage}[t]{0.17\columnwidth}\raggedleft
0.50\strut
\end{minipage} & \begin{minipage}[t]{0.09\columnwidth}\raggedleft
1\strut
\end{minipage}\tabularnewline
\bottomrule
\end{longtable}

\begin{figure}
\centering
\includegraphics[width=6.0in,height=\textheight]{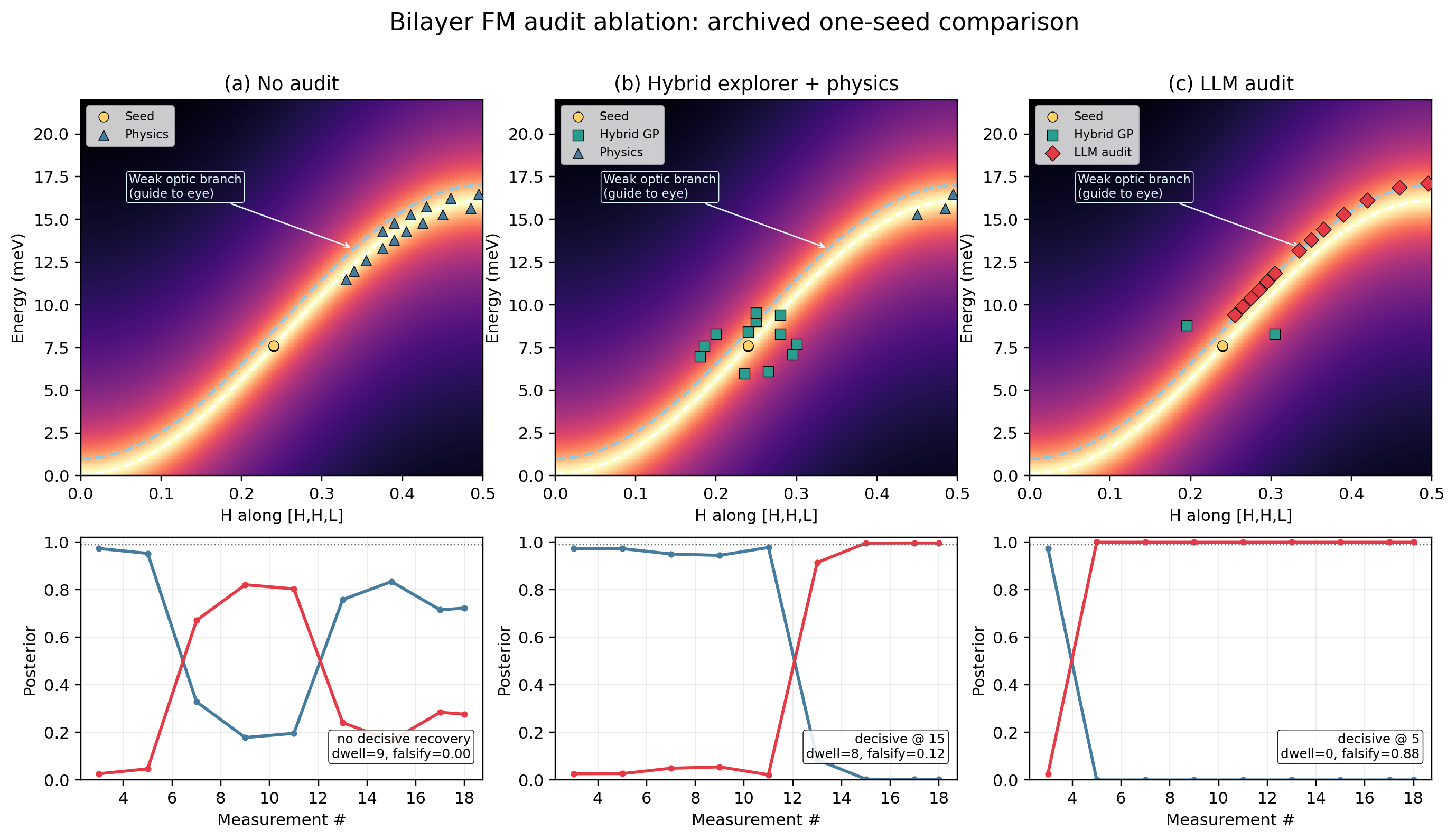}
\caption{Bilayer ferromagnet audit ablation. Top row: executed measurements over the true bilayer intensity map, with dashed guides to the bright acoustic branch and the weak optic branch (labeled as a guide to the eye because its spectral weight is intentionally small). Bottom row: reconstructed posteriors for the monolayer (\(M_A\)) and bilayer (\(M_B\)) models. In panel (a), None uses physics refinement only and eventually recovers, but only after wrong-leader dwell and later decisive selection. In panel (b), Hybrid toggles between bare Log-GP remapping and physics refinement via a deterministic rule and reaches decisive correct selection much earlier. In panel (c), the LLM overseer operates over the same mode choices and may add a bounded number of menu-selected falsification probes (red diamonds). The deterministic Max-disagreement variant reaches decisive correct selection earliest in this comparison (Table 5), while Hybrid and LLM match each other at 8 measurements.}
\end{figure}

The bilayer comparison confirms the pattern from the ghost benchmark:
the falsification channel is the active ingredient. The refinement-only
baseline recovers more slowly and incurs measurable wrong-leader dwell.
All three falsification-capable policies---Hybrid, Max-disagreement, and
LLM---remove that dwell, and the top-two Max-disagreement rule reaches
decisive correct selection earliest. A five-seed rerun reported in
Supplementary Note S5.2 preserves this conclusion: in both the
ghost-optic and bilayer ablations, the LLM and Max-disagreement policies
have identical median decisive times and zero median wrong-leader dwell.
The LLM allocates a large fraction of explicit falsification batches,
but this does not translate into a decisive-time advantage in the
current two-model benchmark.

\hypertarget{scope-and-interpretation}{%
\subsubsection{5.4 Scope and
interpretation}\label{scope-and-interpretation}}

The ghost-optic and bilayer ablations establish a clear result:
\textbf{constrained falsification channels mitigate posterior lock-in}.
In both benchmarks, every policy that explicitly targets falsification
regions---whether a simple top-two Max-disagreement rule, a
threshold-based Hybrid router, or an LLM committee---eliminates
wrong-leader dwell and reaches decisive correct selection far earlier
than refinement-only or undirected-exploration baselines. In these
two-model settings the top-two Max-disagreement rule achieves the same
outcome as the LLM under identical constraints, confirming that the
active ingredient is the falsification principle rather than the
specific implementation; the next paragraph notes where this parity no
longer holds.

This parity is expected in the present two-model benchmarks, where the
decisive falsifier always involves the current top two candidates and is
therefore visible to a top-two disagreement heuristic. However, top-two
disagreement is not universally sufficient. Supplementary Note S5.3
presents a targeted multi-model stress test in which the decisive
falsifier separates the current leader from a lower-ranked model rather
than from the runner-up. In that setting, top-two Max-disagreement is
structurally blind to the critical probe, while a broader falsification
policy (and the LLM) both select it correctly.

The more important question is whether hand-designed heuristics scale to
the diversity of falsification geometries that arise across different
Hamiltonian families. The Max-disagreement rule required choosing a
specific scoring function (top-two intensity difference), a ranking
priority chain, and an injection protocol; when the geometry changed in
the multi-model trap, a different rule (Max-disagreement-all) was
needed. Each new failure mode geometry potentially requires a new
bespoke heuristic. The LLM, by contrast, operates from a
\emph{natural-language description} of the current ambiguity --- ``a
weak optic branch may be missing'' or ``the gap region is
under-sampled'' --- and produces reasonable audit actions from the same
constrained interface without per-problem engineering. For a system
designed to handle unknown Hamiltonians, which is the premise of the
hybrid workflow, this generality across problem descriptions is a
practical advantage that becomes more significant as the Hamiltonian
family grows.

We also note that the posterior evolution in the pilot closed-loop run
(Figure 10) is sensitive to the choice of prior weights: equal priors
can reverse the M4/M2 ranking at intermediate measurement counts
(Supplementary Note S5.4). The chemically informed prior accelerates
discrimination but also shapes it, and this sensitivity should be kept
in mind when interpreting the pilot results.

In summary: algorithmic myopia is real, and constrained falsification
channels fix it robustly across every implementation tested. In
two-model settings a simple deterministic rule suffices; in the
multi-model stress test, top-two disagreement fails structurally while
the LLM selects the correct probe from the same interface it used for
the simpler benchmarks. That consistency across three different
falsification geometries --- ghost-optic, bilayer, and multi-model trap
--- without any per-problem re-engineering is the concrete demonstration
of the generality argument. Whether that generality translates into
systematic performance advantages across broader Hamiltonian families is
an empirical question that the present analytic benchmarks begin to
address but do not exhaust.

\begin{center}\rule{0.5\linewidth}{0.5pt}\end{center}

\hypertarget{discussion-and-outlook}{%
\subsection{6. Discussion and outlook}\label{discussion-and-outlook}}

\hypertarget{hypothesis-generation-from-structure}{%
\subsubsection{6.1 Hypothesis generation from
structure}\label{hypothesis-generation-from-structure}}

The present workflow still assumes a candidate-model library. In
practice, experimentalists often construct that library by inspecting
the crystal structure, identifying exchange paths, and using
Goodenough--Kanamori--Anderson heuristics to propose plausible
Hamiltonians.\textsuperscript{32--34} Figure 13 illustrates the current
hypothesis-generation blueprint used to seed the candidate list, now
with explicit periodic-image path enumeration and an orbital-occupancy
lookup for the Goodenough-Kanamori sign heuristic.

\begin{figure}
\centering
\includegraphics[width=6in,height=\textheight]{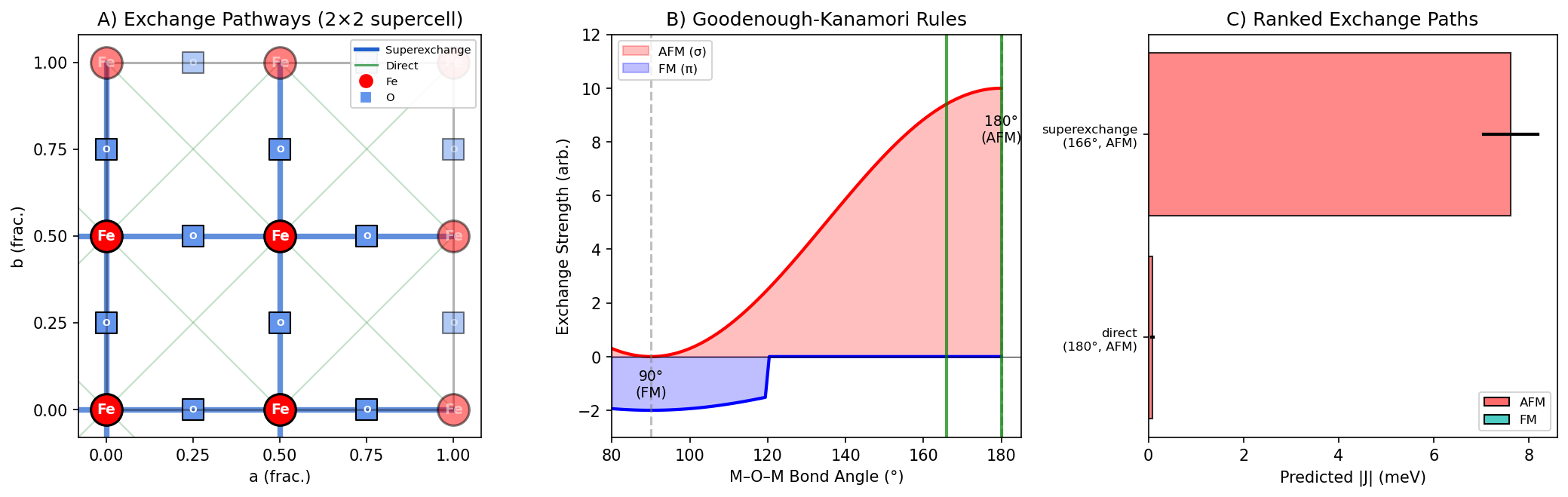}
\caption{Structure-derived hypothesis generation via exchange-path analysis and orbital-aware Goodenough--Kanamori heuristics. Panel A shows the periodic exchange pathways in a 2×2 supercell view, panel B summarizes the angle-based AFM/FM tendencies, and panel C ranks the resulting exchange channels by predicted strength.}
\end{figure}

Automating this step---so that TAS-AI can generate its own candidate
library from a CIF file---is the natural upstream extension of the
workflow. Graph neural network surrogates such as CGCNN, CHGNet, and
related materials models provide a plausible long-term
route,\textsuperscript{35--37} but the present paper focuses on the
downstream discrimination loop once the candidate library exists.

\hypertarget{benchmark-scope-hardware-deployment-and-inference-limitations}{%
\subsubsection{6.2 Benchmark scope, hardware deployment, and inference
limitations}\label{benchmark-scope-hardware-deployment-and-inference-limitations}}

The benchmarks in this paper use square-lattice models with at most
three exchange parameters. This choice is deliberate: simple analytic
models allow the control architecture --- the hybrid handoff, the myopia
diagnosis, and the falsification channel --- to be tested in isolation
without confounding from backend complexity. The claims supported by
these benchmarks are architectural (that staged control outperforms
monolithic control, that falsification channels fix posterior lock-in)
rather than claims about the generality of the physics models.

Establishing physics generality requires broader Hamiltonian families.
The natural next benchmarks would include multi-branch dispersions where
acoustic and optic modes cross or hybridize, anisotropic lattices with
Dzyaloshinskii--Moriya interactions that tilt the dispersion
asymmetrically, and frustrated systems (e.g., triangular or pyrochlore
lattices) where multiple competing ground states produce closely spaced
model candidates. In such settings the Laplace approximation may become
less reliable, the candidate model space will grow, and the
falsification geometry will involve more than two live hypotheses ---
precisely the regime where the multi-model stress test (Supplementary
Note S5.3) suggests richer audit strategies are needed. TAS-AI already
supports PySpinW and Sunny.jl backends that can handle these systems;
the missing ingredient is the benchmark infrastructure and compute time
to run controlled comparisons, not an architectural change to the
controller.

The present work validates the TAS-AI architecture in a high-fidelity
digital twin that includes Cooper--Nathans-derived broadening, realistic
kinematics, motion costs, and counting noise. This digital-twin approach
is deliberate: it allows systematic benchmarking across controlled
scenarios, reproducible ablation studies, and rapid iteration on the
control architecture --- none of which would be feasible during scarce
beam time on a live instrument. Live deployment remains the essential
next step, but the architectural and algorithmic conclusions drawn here
do not depend on it. The inference loop also relies on fast local
approximations for tractability. Full resolution-aware MCMC with complex
backends such as Sunny or PySpinW remains an important next step,
especially for larger Hamiltonian families and non-Gaussian posteriors.

The path to live deployment is architecturally straightforward. On a
real TAS instrument, hardware safety interlocks and collision avoidance
are handled by the instrument protection system, not by the experiment
control software; TAS-AI's kinematic constraints (accessible
reciprocal-space window, energy limits, motor-speed bounds) are the
software-level analogue and are already implemented. Motor backlash
affects the optimal direction of constant-\(Q\) or constant-\(E\) scans
and would need to be incorporated into the motion-cost model for
production use, but does not change the control architecture. The main
integration requirements are asynchronous I/O between the planner and
the instrument control system (NICE, SICS, or equivalent), latency
tolerance for the sub-second planning loop, and operator visibility into
the autonomous queue. As the NIST Center for Neutron Research (NCNR)
returns to scientific operations, validating the hybrid workflow on live
instruments will be essential to confirming the practical gains
demonstrated here.

\begin{center}\rule{0.5\linewidth}{0.5pt}\end{center}

\hypertarget{conclusions}{%
\subsection{7. Conclusions}\label{conclusions}}

Autonomous TAS is best understood as a \textbf{hybrid sequential design
problem} rather than as a search for a single universally optimal
acquisition rule. Detection, inference, and refinement are different
tasks, and they reward different controllers.

This paper establishes four main results.

\begin{enumerate}
\def\labelenumi{\arabic{enumi}.}
\tightlist
\item
  \textbf{Agnostic discovery is the correct front end for unknown
  spectra.} On blind global reconstruction metrics, agnostic methods
  outperform physics-only planning, showing why a hybrid workflow is
  necessary rather than optional.
\item
  \textbf{Physics-informed planning becomes valuable once structure is
  present.} In controlled benchmarks, TAS-AI rapidly contracts parameter
  uncertainty and reaches decisive in-loop model discrimination using
  AIC-derived evidence ratios as a practical real-time proxy.
\item
  \textbf{Time-aware planning matters experimentally.} Motion-aware
  acquisition and MCTS batch sequencing translate information gain into
  real wall-clock savings on an instrument where motor travel is often
  comparable to count time.
\item
  \textbf{Constrained falsification channels mitigate posterior
  lock-in.} In controlled ablations, every policy that explicitly
  targets falsification regions---including a simple max-disagreement
  rule---eliminates wrong-leader dwell and reaches decisive model
  selection far earlier than refinement-only baselines. The active
  ingredient is the falsification principle, not the specific
  implementation. A targeted multi-model stress test identifies a regime
  where top-two heuristics are structurally blind, and the LLM committee
  offers generality across diverse problem descriptions without
  per-problem engineering---an architectural advantage for systems
  facing unknown Hamiltonians.
\end{enumerate}

As autonomous capabilities expand across neutron facilities and other
user instruments, the broader design principle is clear:
\textbf{agnostic controllers excel at discovery; physics-informed
controllers excel at inference; and audit layers are needed when greedy
utilities become self-confirming.} TAS-AI provides an open-source
foundation for exploring that architecture in spin-wave spectroscopy and
beyond.

\begin{center}\rule{0.5\linewidth}{0.5pt}\end{center}

\hypertarget{data-availability}{%
\subsection{Data Availability}\label{data-availability}}

The TAS-AI library --- physics backends (including the analytic
\texttt{SquareLatticeAFM} and \texttt{SquareFMBilayer} models used in
the closed-loop pilots), the acquisition, instrument-resolution, and
MCTS modules, and the core audit interface --- is available at
\texttt{https://github.com/usnistgov/tasai}. The manuscript source,
figure-generation scripts, closed-loop drivers
(\texttt{toy\_closed\_loop.py}, \texttt{run\_audit\_ablation.py}, and
associated overseer wiring), benchmark summaries, prior specifications,
and paper-specific provenance artifacts are available at
\texttt{https://github.com/usnistgov/paper-tasai}. Prompt artifacts used
for the pilot audit layer are described in the Supplementary
Information.

\hypertarget{author-contributions}{%
\subsection{Author Contributions}\label{author-contributions}}

W.R. conceived the project, developed the methodology, implemented the
software, and wrote the manuscript.

\hypertarget{conflicts-of-interest}{%
\subsection{Conflicts of Interest}\label{conflicts-of-interest}}

There are no conflicts to declare.

\hypertarget{acknowledgements}{%
\subsection{Acknowledgements}\label{acknowledgements}}

This work used the Texas Advanced Computing Center (TACC) at The
University of Texas at Austin for benchmark sweeps on Stampede3 (project
CDA24014). During the preparation of this work, the author used
generative AI tools for language polishing, editing, and assistance with
rewriting portions of the manuscript. After using these tools, the
author reviewed and edited the content as needed and takes full
responsibility for the publication's content.

\hypertarget{references}{%
\subsection{References}\label{references}}

Certain commercial equipment, instruments, or materials (or suppliers,
or software, \ldots) are identified in this paper to foster
understanding. Such identification does not imply recommendation or
endorsement by the National Institute of Standards and Technology, nor
does it imply that the materials or equipment identified are necessarily
the best available for the purpose.

\hypertarget{refs}{}
\begin{cslreferences}
\leavevmode\hypertarget{ref-ratcliff2016review}{}%
1 W. Ratcliff, J. W. Lynn, V. Kiryukhin, P. Jain and M. R. Fitzsimmons,
Magnetic structures and dynamics of multiferroic systems obtained with
neutron scattering, \emph{npj Quantum Materials}, 2016, \textbf{1},
16003.

\leavevmode\hypertarget{ref-senff2007tbmno3}{}%
2 D. Senff, P. Link, K. Hradil, A. Hiess, L. P. Regnault, Y. Sidis, N.
Aliouane, D. N. Argyriou and M. Braden, Magnetic excitations in
multiferroic TbMnO\(_3\): Evidence for a hybridized soft mode,
\emph{Physical Review Letters}, 2007, \textbf{98}, 137206.

\leavevmode\hypertarget{ref-jeong2012bifeo3}{}%
3 J. Jeong, E. A. Goremychkin, T. Guidi, K. Nakajima, G. S. Jeon, S.-A.
Kim, S. Furukawa, Y. B. Kim, S. Lee, V. Kiryukhin, S.-W. Cheong and
J.-G. Park, Spin wave measurements over the full Brillouin zone of
multiferroic BiFeO\(_3\), \emph{Physical Review Letters}, 2012,
\textbf{108}, 077202.

\leavevmode\hypertarget{ref-matsuda2012bifeo3}{}%
4 M. Matsuda, R. S. Fishman, T. Hong, C. H. Lee, T. Ushiyama, Y.
Yanagisawa, Y. Tomioka and T. Ito, Magnetic dispersion and anisotropy in
multiferroic BiFeO\(_3\), \emph{Physical Review Letters}, 2012,
\textbf{109}, 067205.

\leavevmode\hypertarget{ref-disseler2015lufeo3}{}%
5 S. M. Disseler, X. Luo, B. Gao, Y. S. Oh, R. Hu, Y. Wang, D. Quintana,
A. Zhang, Q. Huang, J. Lau, R. Paul, J. W. Lynn, S.-W. Cheong and W.
Ratcliff, Multiferroicity in doped hexagonal LuFeO\(_3\), \emph{Physical
Review B}, 2015, \textbf{92}, 054435.

\leavevmode\hypertarget{ref-coldea2001la2cuo4}{}%
6 R. Coldea, S. M. Hayden, G. Aeppli, T. G. Perring, C. D. Frost, T. E.
Mason, S.-W. Cheong and Z. Fisk, Spin waves and electronic interactions
in La\(_2\)CuO\(_4\), \emph{Physical Review Letters}, 2001, \textbf{86},
5377--5380.

\leavevmode\hypertarget{ref-brockhouse1995}{}%
7 B. N. Brockhouse, Slow neutron spectroscopy and the grand atlas of the
physical world, \emph{Reviews of Modern Physics}, 1995, \textbf{67},
735--751.

\leavevmode\hypertarget{ref-jones1998ei}{}%
8 D. R. Jones, M. Schonlau and W. J. Welch, Efficient global
optimization of expensive black-box functions, \emph{Journal of Global
Optimization}, 1998, \textbf{13}, 455--492.

\leavevmode\hypertarget{ref-srinivas2010ucb}{}%
9 N. Srinivas, A. Krause, S. M. Kakade and M. Seeger, in
\emph{Proceedings of the 27th international conference on machine
learning (ICML)}, 2010, pp. 1015--1022.

\leavevmode\hypertarget{ref-hippalgaonkar2023}{}%
10 K. Hippalgaonkar, Q. Li, X. Wang, J. W. Fisher, J. Kirkpatrick and T.
Buonassisi, Knowledge-integrated machine learning for materials: Lessons
from gameplaying and robotics, \emph{Nature Reviews Materials}, 2023,
\textbf{8}, 241--260.

\leavevmode\hypertarget{ref-abolhasani2023}{}%
11 M. Abolhasani and E. Kumacheva, The rise of self-driving labs in
chemical and materials sciences, \emph{Nature Synthesis}, 2023,
\textbf{2}, 483--492.

\leavevmode\hypertarget{ref-burger2020}{}%
12 B. Burger, P. M. Maffettone, V. V. Gusev, C. M. Aitchison, Y. Bai, X.
Wang, X. Li, B. M. Alston, B. Li, R. Clowes, N. Rankin, B. Harris, R. S.
Sprick and A. I. Cooper, A mobile robotic chemist, \emph{Nature}, 2020,
\textbf{583}, 237--241.

\leavevmode\hypertarget{ref-hwang2026korea}{}%
13 J. Hwang, S. Kim, S. Lim, J. Kim, S. Lee, S. Min, J. Song, J. Lim, S.
Hong, J.-H. Hwang, Y.-S. Choi, D.-H. Seo, S. S. Han, K. Kim, S.-H. Yoo,
J. Shin, J. W. Choi, J. Nam, J. Park, J. Ryu and Y. Jung, Self-driving
laboratories in korea: A new era of autonomous discovery, \emph{Digital
Discovery},
DOI:\href{https://doi.org/10.1039/D6DD00024J}{10.1039/D6DD00024J}.

\leavevmode\hypertarget{ref-kusne2020}{}%
14 A. G. Kusne, H. Yu, C. Wu, H. Zhang, J. Hattrick-Simpers, B. DeCost,
S. Sarker, C. Oses, C. Toher, S. Curtarolo, A. V. Davydov, R. Agarwal,
L. A. Bendersky, M. Li, A. Mehta and I. Takeuchi, On-the-fly closed-loop
materials discovery via bayesian active learning, \emph{Nature
Communications}, 2020, \textbf{11}, 5966.

\leavevmode\hypertarget{ref-ament2021}{}%
15 S. Ament, M. Amsler, D. R. Sutherland, M.-C. Chang, D. Guevarra, A.
B. Connolly, J. M. Gregoire, M. O. Thompson, C. P. Gomes and R. B. van
Dover, Autonomous materials synthesis via hierarchical active learning
of nonequilibrium phase diagrams, \emph{Science Advances}, 2021,
\textbf{7}, eabg4930.

\leavevmode\hypertarget{ref-noack2019}{}%
16 M. M. Noack, K. G. Yager, M. Fukuto, G. S. Doerk, R. Li and J. A.
Sethian, A Kriging-based approach to autonomous experimentation with
applications to x-ray scattering, \emph{Scientific Reports}, 2019,
\textbf{9}, 11809.

\leavevmode\hypertarget{ref-noack2021}{}%
17 M. M. Noack, P. H. Zwart, D. M. Ushizima, M. Fukuto, K. G. Yager, K.
C. Elbert, C. B. Murray, A. Stein, G. S. Doerk, E. H. R. Tsai, R. Li, G.
Freychet, M. Zhernenkov, H.-Y. N. Holman, S. Lee, L. Chen, E. Rotenberg,
T. Weber, Y. Le Goc, M. Boehm, P. Steffens, P. Mutti and J. A. Sethian,
Gaussian processes for autonomous data acquisition at large-scale
synchrotron and neutron facilities, \emph{Nature Reviews Physics}, 2021,
\textbf{3}, 685--697.

\leavevmode\hypertarget{ref-teixeiraparente2023natcomms}{}%
18 M. Teixeira Parente, G. Brandl, C. Franz, U. Stuhr, M. Ganeva and A.
Schneidewind, Active learning-assisted neutron spectroscopy with
log-Gaussian processes, \emph{Nature Communications}, 2023, \textbf{14},
2246.

\leavevmode\hypertarget{ref-hoogerheide2024autorefl}{}%
19 D. P. Hoogerheide and F. Heinrich, AutoRefl: Active learning in
neutron reflectometry for fast data acquisition, \emph{Journal of
Applied Crystallography}, 2024, \textbf{57}, 1192--1204.

\leavevmode\hypertarget{ref-huarcaya2026autonse}{}%
20 R. Huarcaya, A. McDannald, W. D. Ratcliff and D. P. Hoogerheide,
Autonomous data collection for the neutron spin echo response function,
\emph{APL Machine Learning},
DOI:\href{https://doi.org/10.1063/5.0312231}{10.1063/5.0312231}.

\leavevmode\hypertarget{ref-mcdannald2022andie}{}%
21 A. McDannald, M. Frontzek, A. T. Savici, M. Doucet, E. E. Rodriguez,
K. Meuse, J. Opsahl-Ong, D. Samarov, I. Takeuchi, W. Ratcliff and A. G.
Kusne, On-the-fly autonomous control of neutron diffraction via
physics-informed Bayesian active learning, \emph{Applied Physics
Reviews}, 2022, \textbf{9}, 021408.

\leavevmode\hypertarget{ref-teixeiraparente2022front}{}%
22 M. Teixeira Parente, A. Schneidewind, G. Brandl, C. Franz, M. Noack,
M. Boehm and M. Ganeva, Benchmarking autonomous scattering experiments
illustrated on TAS, \emph{Frontiers in Materials}, 2022, \textbf{8},
772014.

\leavevmode\hypertarget{ref-akaike1974}{}%
23 H. Akaike, A new look at the statistical model identification,
\emph{IEEE Transactions on Automatic Control}, 1974, \textbf{19},
716--723.

\leavevmode\hypertarget{ref-schwarz1978}{}%
24 G. Schwarz, Estimating the dimension of a model, \emph{The Annals of
Statistics}, 1978, \textbf{6}, 461--464.

\leavevmode\hypertarget{ref-kass1995}{}%
25 R. E. Kass and A. E. Raftery, Bayes factors, \emph{Journal of the
American Statistical Association}, 1995, \textbf{90}, 773--795.

\leavevmode\hypertarget{ref-coopernathans1967a}{}%
26 M. J. Cooper and R. Nathans, The resolution function in neutron
diffractometry. I. The resolution function of a neutron diffractometer
and its application to phonon measurements, \emph{Acta
Crystallographica}, 1967, \textbf{23}, 357--367.

\leavevmode\hypertarget{ref-coopernathans1968b}{}%
27 M. J. Cooper and R. Nathans, The resolution function in neutron
diffractometry. III. Experimental determination and properties of the
`elastic two-crystal' resolution function, \emph{Acta Crystallographica
Section A}, 1968, \textbf{24}, 619--624.

\leavevmode\hypertarget{ref-toth2015}{}%
28 S. Toth and B. Lake, Linear spin wave theory for single-q
incommensurate magnetic structures, \emph{Journal of Physics: Condensed
Matter}, 2015, \textbf{27}, 166002.

\leavevmode\hypertarget{ref-dahlbom2025sunny}{}%
29 D. Dahlbom, H. Zhang, C. Miles, S. Quinn, A. Niraula, B. Thipe, M.
Wilson, S. Matin, H. Mankad, S. Hahn, D. Pajerowski, S. Johnston, Z.
Wang, H. Lane, Y. W. Li, X. Bai, M. Mourigal, C. D. Batista and K.
Barros, Sunny.jl: A julia package for spin dynamics, \emph{Journal of
Open Source Software}, 2025, \textbf{10}, 8138.

\leavevmode\hypertarget{ref-kocsis2006uct}{}%
30 L. Kocsis and C. Szepesvári, in \emph{Machine learning: ECML 2006},
Springer, Berlin, Heidelberg, 2006, vol. 4212, pp. 282--293.

\leavevmode\hypertarget{ref-browne2012mcts}{}%
31 C. B. Browne, E. Powley, D. Whitehouse, S. M. Lucas, P. I. Cowling,
P. Rohlfshagen, S. Tavener, D. Perez, S. Samothrakis and S. Colton, A
survey of Monte Carlo tree search methods, \emph{IEEE Transactions on
Computational Intelligence and AI in Games}, 2012, \textbf{4}, 1--43.

\leavevmode\hypertarget{ref-goodenough1955}{}%
32 J. B. Goodenough, Theory of the role of covalence in the
perovskite-type manganites {[}La, M(II){]}MnO\(_3\), \emph{Physical
Review}, 1955, \textbf{100}, 564--573.

\leavevmode\hypertarget{ref-kanamori1959}{}%
33 J. Kanamori, Superexchange interaction and symmetry properties of
electron orbitals, \emph{Journal of Physics and Chemistry of Solids},
1959, \textbf{10}, 87--98.

\leavevmode\hypertarget{ref-anderson1950}{}%
34 P. W. Anderson, Antiferromagnetism. Theory of superexchange
interaction, \emph{Physical Review}, 1950, \textbf{79}, 350--356.

\leavevmode\hypertarget{ref-xie2018cgcnn}{}%
35 T. Xie and J. C. Grossman, Crystal graph convolutional neural
networks for an accurate and interpretable prediction of material
properties, \emph{Physical Review Letters}, 2018, \textbf{120}, 145301.

\leavevmode\hypertarget{ref-deng2023chgnet}{}%
36 B. Deng, P. Zhong, K. Jun, J. Riebesell, K. Han, C. J. Bartel and G.
Ceder, CHGNet as a pretrained universal neural network potential for
charge-informed atomistic modelling, \emph{Nature Machine Intelligence},
2023, \textbf{5}, 1031--1041.

\leavevmode\hypertarget{ref-chen2022m3gnet}{}%
37 C. Chen and S. P. Ong, A universal graph deep learning interatomic
potential for the periodic table, \emph{Nature Computational Science},
2022, \textbf{2}, 718--728.
\end{cslreferences}

\clearpage
\appendix
\setcounter{secnumdepth}{3}
\renewcommand{\thesubsection}{S\arabic{subsection}}
\renewcommand{\thesubsubsection}{S\arabic{subsection}.\arabic{subsubsection}}
\renewcommand{\thefigure}{S\arabic{figure}}
\renewcommand{\thetable}{S\arabic{table}}
\renewcommand{\theequation}{S\arabic{equation}}
\setcounter{figure}{0}
\setcounter{table}{0}
\setcounter{equation}{0}
\setcounter{subsection}{0}
\setcounter{subsubsection}{0}
\section*{Supplementary Information}
\addcontentsline{toc}{section}{Supplementary Information}

\hypertarget{enhanced-log-gp-the-1d-taper-and-linear-intensity-variance-weighting}{%
\subsection{Enhanced Log-GP: the 1D taper and linear-intensity variance
weighting}\label{enhanced-log-gp-the-1d-taper-and-linear-intensity-variance-weighting}}

The underlying Log-GP reconstruction idea is due to Teixeira Parente
\emph{et al.} and the JCNS neutron active-learning work cited in the
main text; it is not introduced in this manuscript. What is specific to
the present implementation is the set of safeguards that stabilize the
Log-GP policy for our TAS benchmark domain.

In the agnostic Log-GP phase, uncertainty sampling in
\textbf{log-intensity space} can over-prioritize the boundaries of the
search domain. GP predictive variance is naturally largest at the edges
of a bounded box, and log-variance treats dim background regions as
comparably valuable to bright signal regions. The resulting acquisition
can become \textbf{edge-locked}, repeatedly sampling high-energy or
high-\(|H|\) boundary points that do not intersect real signal support.

We mitigate this failure mode with two complementary changes.

\begin{enumerate}
\def\labelenumi{\arabic{enumi}.}
\tightlist
\item
  \textbf{Linear-intensity variance weighting.} Rather than ranking
  candidates by raw log-space variance alone, the acquisition is
  weighted by a linear-space variance proxy so that dim background
  regions are not treated as equally valuable as bright signal regions.
  When the surrogate exposes log-space mean and variance
  \((\mu, \sigma^2)\) directly, this can be written as
\end{enumerate}

\[
\mathrm{Var}(I) = \left(e^{\sigma^2}-1\right)e^{2\mu+\sigma^2},
\]

where \(\mu\) and \(\sigma^2\) are the GP posterior mean and variance in
log-intensity space and \(\mathrm{Var}(I)\) is the corresponding
approximate variance in linear intensity units. In the current codebase,
the live path may instead use the backend's directly exposed posterior
standard deviation in linear space; the common design principle is that
acquisition is ranked in linear-intensity variance units rather than raw
log-space variance.

\begin{enumerate}
\def\labelenumi{\arabic{enumi}.}
\setcounter{enumi}{1}
\tightlist
\item
  \textbf{A 1D cosine taper in energy.} We apply a soft window in \(E\)
  that smoothly downweights the outer 10\% of the energy domain while
  leaving the interior nearly unchanged.
\end{enumerate}

A stronger 2D taper in both \(E\) and \(H\) further suppresses edge
selection, but in this model it can over-penalize low-\(|H|\) regions
where the dispersion is strongest. We therefore retain the energy-only
taper in the reported benchmarks.

\begin{figure}
\hypertarget{fig:loggp-taper}{%
\centering
\includegraphics[width=0.9\textwidth,height=\textheight]{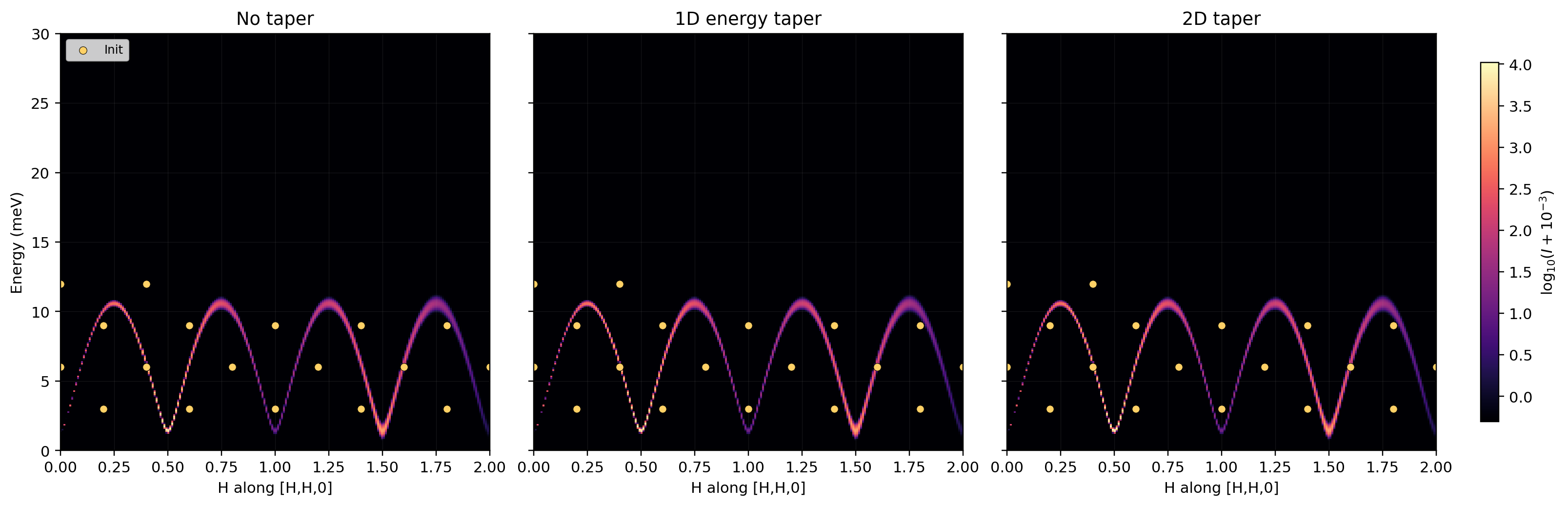}
\caption{Effect of linear-intensity variance weighting and boundary
tapering on Log-GP active selection. Left: no taper (edge locked).
Middle: 1D energy taper used in this work. Right: stronger 2D taper,
which was tested but suppresses low-\(|H|\) ridge coverage in this
model.}\label{fig:loggp-taper}
}
\end{figure}

\hypertarget{benchmark-runtime-accounting}{%
\subsection{Benchmark runtime
accounting}\label{benchmark-runtime-accounting}}

Detailed benchmark provenance has been moved out of the SI and into the
reproducibility/referee material:

\begin{itemize}
\tightlist
\item
  \texttt{REPRODUCIBILITY.md}
\item
  \texttt{REVIEWER\_GUIDE.md}
\item
  \texttt{paper/data/README.md}
\item
  \texttt{paper/data/table2\_provenance.md}
\end{itemize}

The controller itself operates at sub-second algorithmic latency, but
benchmark sweeps incur additional end-to-end digital-twin overhead.
Table S1 reports the canonical planner-side quantity preserved in the
benchmark JSON files: \texttt{mean\_time\_per\_suggestion}. This is the
reproducible per-suggestion compute cost of the benchmark harness
itself. We do not report a separate elapsed-per-run column because those
wall-clock values depend on execution environment and are not preserved
consistently in the final benchmark artifacts.

\emph{Table S1.} Mean planner-side compute time per simulated suggestion
from the canonical benchmark JSON artifacts. Internal archive keys such
as \texttt{faira\_*} and \texttt{fairpf\_*} are omitted here because
they are implementation-facing provenance labels rather than scientific
benchmark categories.

\begin{longtable}[]{@{}llrr@{}}
\toprule
\textbf{Benchmark family} & \textbf{Method} & \textbf{Mean time per
suggestion (ms)} & \textbf{Std. dev. (ms)}\tabularnewline
\midrule
\endhead
Analytic & Grid & 0.00028 & 0.00001\tabularnewline
Analytic & Random & 0.00959 & 0.00034\tabularnewline
Analytic & Enhanced Log-GP & 847.8 & 58.4\tabularnewline
Analytic & TAS-AI (physics) & 15.98 & 0.23\tabularnewline
PySpinW+CN & Grid & 0.00081 & 0.00002\tabularnewline
PySpinW+CN & Random & 0.01658 & 0.00032\tabularnewline
PySpinW+CN & Enhanced Log-GP & 965.9 & 56.6\tabularnewline
PySpinW+CN & TAS-AI (physics) & 16.03 & 0.23\tabularnewline
\bottomrule
\end{longtable}

These values come directly from the archived benchmark outputs and are
therefore reproducible from the current repository state. They should be
interpreted as planner-side digital-twin cost, not beamtime or
wall-clock queueing cost.

\hypertarget{blind-reconstruction-metric-and-threshold}{%
\subsubsection{Blind-reconstruction metric and
threshold}\label{blind-reconstruction-metric-and-threshold}}

The blind benchmark figures and Table 1 use a common reconstruction
metric defined on a fixed reference grid in \((H,E)\) space:

\[
\varepsilon_{\mathrm{recon}}
=
\frac{\sum_i \left|I_i^{\mathrm{pred}}-I_i^{\mathrm{true}}\right|\,I_i^{\mathrm{true}}}
{\sum_i \left(I_i^{\mathrm{true}}\right)^2}.
\]

Here \(I_i^{\mathrm{true}}\) is the ground-truth intensity at
reference-grid point \(i\), and \(I_i^{\mathrm{pred}}\) is the
reconstructed intensity inferred from the method's raw measurements. In
the current benchmark implementation, \(I_i^{\mathrm{pred}}\) is
obtained from inverse-distance interpolation over the raw observed
intensities rather than from each method's internal surrogate state.
This makes the score an acquisition-quality metric rather than a
benchmark of surrogate-specific reconstruction machinery. The canonical
implementation is \texttt{compute\_reconstruction\_error()} in
\texttt{tasai/examples/benchmark\_jcns.py}.

For the analytic benchmark families, a run is counted as successful when
\(\varepsilon_{\mathrm{recon}} \le 0.20\) within the fixed budget. For
the corrected PySpinW+Cooper-Nathans rows, no method reaches that
threshold within \(N=300\), so the informative quantity is the final
error at budget rather than the median measurements-to-threshold.

\hypertarget{fixed-count-time-and-mcts-settings}{%
\subsubsection{Fixed count time and MCTS
settings}\label{fixed-count-time-and-mcts-settings}}

In the current time-aware refinement study, count time is \textbf{not}
optimized jointly with location. The refinement benchmark uses a fixed
dwell time of 10 s per measurement and optimizes only the
location-dependent information-rate objective. This is why the
wall-clock gains in Figure 5 come from route choice and information
density rather than from adaptive dwell-time allocation.

When the optional MCTS batch planner is used, the current core defaults
are:

\begin{itemize}
\tightlist
\item
  \texttt{n\_simulations\ =\ 100}
\item
  \texttt{exploration\_constant\ =\ 1.41}
\item
  \texttt{n\_candidates\ =\ 20}
\item
  \texttt{rollout\_depth\ =\ 3}
\item
  \texttt{discount\_factor\ =\ 0.95}
\item
  \texttt{max\_depth\ =} requested batch size
\end{itemize}

These values are taken directly from the active implementation in
\texttt{tasai/core/mcts.py}. They were chosen as practical short-horizon
settings for motion-aware batched planning rather than as a separately
optimized benchmark target.

\hypertarget{additional-remarks-on-stopping-local-approximation-and-escalation}{%
\subsection{Additional remarks on stopping, local approximation, and
escalation}\label{additional-remarks-on-stopping-local-approximation-and-escalation}}

Inside the physics-informed loop, TAS-AI uses fast multi-start local
fits and a Laplace/Levenberg--Marquardt covariance approximation because
the planner must evaluate many candidates in real time. These
approximations are appropriate when the posterior is locally unimodal
and the candidate model family is already close to the truth, but they
can under-represent uncertainty in strongly multimodal settings.

To guard against that failure mode, the code uses simple escalation
triggers. In the controlled single-branch tests, reduced \(\chi^2\)
values that remain high (for example, above approximately 5) or
parameter estimates that repeatedly hit bounds trigger a heavier
posterior stage. In current practice, that heavier stage is reserved for
batch boundaries or offline validation rather than every in-loop update.

These escalation rules are part of the reason the manuscript
distinguishes three operating regimes: agnostic discovery,
physics-informed inference, and strategic audit. Each regime uses a
different approximation budget and a different notion of what
constitutes ``useful'' information.

\hypertarget{small-seed-robustness-checks-for-refinement-and-discrimination}{%
\subsubsection{Small-seed robustness checks for refinement and
discrimination}\label{small-seed-robustness-checks-for-refinement-and-discrimination}}

To provide a first robustness check beyond the representative runs shown
in the main text, we ran a small three-seed sensitivity sweep for the
controlled time-aware refinement benchmark of Figure 5 and for the
simple NN-vs-\(J_1\)-\(J_2\) discrimination benchmark of Figure 6. These
results are archived in
\texttt{paper/data/reviewer\_sensitivity\_20260403.json}.

For the Figure 5 refinement setup, the ranking across three seeds is:

\begin{itemize}
\tightlist
\item
  \texttt{tas\_ai} succeeds in 3/3 runs, with median convergence at 11
  measurements and median convergence time 225 s.
\item
  \texttt{random} also succeeds in 3/3 runs, with the same median
  measurement count but median convergence time 413 s.
\item
  \texttt{grid} does not reach the threshold in any of the 3 runs within
  the same budget.
\end{itemize}

The multi-seed result therefore supports a \textbf{wall-clock} advantage
for TAS-AI rather than a measurement-count advantage: the motion-aware
planner is consistently faster in elapsed time even when the number of
measurements needed is similar to the best random runs.

For the simple NN-vs-\(J_1\)-\(J_2\) discrimination benchmark, the
result is stable across the tested seeds: all 3/3 runs reach decisive
evidence after the initial six-point seed set. This is consistent with
the main-text claim that once the model family is appropriate and the
discriminating region is already sampled, the in-loop model-selection
signal is very strong.

\hypertarget{sensitivity-to-the-exploration-exponent}{%
\subsubsection{Sensitivity to the exploration
exponent}\label{sensitivity-to-the-exploration-exponent}}

The motion-aware refinement policy uses an empirical exploration
exponent \(\eta\) (denoted \(\gamma\) in the main text) in the
information-rate score. A small three-seed sweep over
\(\eta \in \{0.5, 0.7, 0.9\}\) gives:

\begin{longtable}[]{@{}rrrrr@{}}
\toprule
\begin{minipage}[b]{0.08\columnwidth}\raggedleft
\textbf{\(\eta\)}\strut
\end{minipage} & \begin{minipage}[b]{0.09\columnwidth}\raggedleft
\textbf{Success}\strut
\end{minipage} & \begin{minipage}[b]{0.29\columnwidth}\raggedleft
\textbf{Median measurements to threshold}\strut
\end{minipage} & \begin{minipage}[b]{0.25\columnwidth}\raggedleft
\textbf{Median convergence time (s)}\strut
\end{minipage} & \begin{minipage}[b]{0.15\columnwidth}\raggedleft
\textbf{Mean final RMS}\strut
\end{minipage}\tabularnewline
\midrule
\endhead
\begin{minipage}[t]{0.08\columnwidth}\raggedleft
0.5\strut
\end{minipage} & \begin{minipage}[t]{0.09\columnwidth}\raggedleft
3/3\strut
\end{minipage} & \begin{minipage}[t]{0.29\columnwidth}\raggedleft
11\strut
\end{minipage} & \begin{minipage}[t]{0.25\columnwidth}\raggedleft
225\strut
\end{minipage} & \begin{minipage}[t]{0.15\columnwidth}\raggedleft
0.0169\strut
\end{minipage}\tabularnewline
\begin{minipage}[t]{0.08\columnwidth}\raggedleft
0.7\strut
\end{minipage} & \begin{minipage}[t]{0.09\columnwidth}\raggedleft
3/3\strut
\end{minipage} & \begin{minipage}[t]{0.29\columnwidth}\raggedleft
11\strut
\end{minipage} & \begin{minipage}[t]{0.25\columnwidth}\raggedleft
225\strut
\end{minipage} & \begin{minipage}[t]{0.15\columnwidth}\raggedleft
0.0104\strut
\end{minipage}\tabularnewline
\begin{minipage}[t]{0.08\columnwidth}\raggedleft
0.9\strut
\end{minipage} & \begin{minipage}[t]{0.09\columnwidth}\raggedleft
3/3\strut
\end{minipage} & \begin{minipage}[t]{0.29\columnwidth}\raggedleft
11\strut
\end{minipage} & \begin{minipage}[t]{0.25\columnwidth}\raggedleft
220\strut
\end{minipage} & \begin{minipage}[t]{0.15\columnwidth}\raggedleft
0.0066\strut
\end{minipage}\tabularnewline
\bottomrule
\end{longtable}

Within this limited sweep, the refinement result is not brittle across
the tested range. The higher value \(\eta=0.9\) is slightly better on
final RMS and slightly faster in median elapsed time, while the current
default \(\eta=0.7\) remains safely inside the stable regime rather than
at a knife-edge optimum.

\hypertarget{aic-versus-waic-in-the-simple-discrimination-test}{%
\subsubsection{AIC versus WAIC in the simple discrimination
test}\label{aic-versus-waic-in-the-simple-discrimination-test}}

The main manuscript uses AIC-derived weights as a pragmatic real-time
model-selection proxy. As a focused check on that choice, we computed an
offline grid-based WAIC comparison for the simple NN-vs-\(J_1\)-\(J_2\)
discrimination setup after 8 measurements. In the tested seed, AIC and
WAIC agree completely on the model ranking: both assign effectively unit
weight to the correct \(J_1\)-\(J_2\) model and negligible weight to the
NN-only alternative.

This does \textbf{not} prove that AIC and WAIC are interchangeable in
all TAS-AI regimes. It does show that, in the simple controlled
discrimination setting corresponding to Figure 6, the manuscript's
AIC-based conclusion is not being driven by a disagreement with this
standard offline predictive criterion.

\hypertarget{empirical-coverage-of-the-laplace-credible-intervals}{%
\subsubsection{Empirical coverage of the Laplace credible
intervals}\label{empirical-coverage-of-the-laplace-credible-intervals}}

To check whether the fast Laplace/Levenberg--Marquardt uncertainty
estimates are actually calibrated in the controlled Figure 5 refinement
setting, we ran a 10-seed coverage calculation on the TAS-AI policy. For
each seed we computed the final local covariance from the numerical
Hessian of the \(\chi^2\) objective at the converged parameter estimate,
formed nominal 90\% marginal intervals for \((J_1,J_2,D)\), and counted
the fraction of seeds in which those intervals contained the true
parameter values. The seed-level results are archived in
\texttt{paper/data/laplace\_coverage\_refinement\_20260415.json}.

\begin{longtable}[]{@{}rrrrr@{}}
\toprule
\begin{minipage}[b]{0.11\columnwidth}\raggedleft
\textbf{Parameter}\strut
\end{minipage} & \begin{minipage}[b]{0.17\columnwidth}\raggedleft
\textbf{Nominal coverage}\strut
\end{minipage} & \begin{minipage}[b]{0.18\columnwidth}\raggedleft
\textbf{Empirical coverage}\strut
\end{minipage} & \begin{minipage}[b]{0.11\columnwidth}\raggedleft
\textbf{Hit count}\strut
\end{minipage} & \begin{minipage}[b]{0.29\columnwidth}\raggedleft
\textbf{Median 90\% interval width (meV)}\strut
\end{minipage}\tabularnewline
\midrule
\endhead
\begin{minipage}[t]{0.11\columnwidth}\raggedleft
\(J_1\)\strut
\end{minipage} & \begin{minipage}[t]{0.17\columnwidth}\raggedleft
0.90\strut
\end{minipage} & \begin{minipage}[t]{0.18\columnwidth}\raggedleft
0.30\strut
\end{minipage} & \begin{minipage}[t]{0.11\columnwidth}\raggedleft
3/10\strut
\end{minipage} & \begin{minipage}[t]{0.29\columnwidth}\raggedleft
0.0171\strut
\end{minipage}\tabularnewline
\begin{minipage}[t]{0.11\columnwidth}\raggedleft
\(J_2\)\strut
\end{minipage} & \begin{minipage}[t]{0.17\columnwidth}\raggedleft
0.90\strut
\end{minipage} & \begin{minipage}[t]{0.18\columnwidth}\raggedleft
0.70\strut
\end{minipage} & \begin{minipage}[t]{0.11\columnwidth}\raggedleft
7/10\strut
\end{minipage} & \begin{minipage}[t]{0.29\columnwidth}\raggedleft
0.0277\strut
\end{minipage}\tabularnewline
\begin{minipage}[t]{0.11\columnwidth}\raggedleft
\(D\)\strut
\end{minipage} & \begin{minipage}[t]{0.17\columnwidth}\raggedleft
0.90\strut
\end{minipage} & \begin{minipage}[t]{0.18\columnwidth}\raggedleft
0.10\strut
\end{minipage} & \begin{minipage}[t]{0.11\columnwidth}\raggedleft
1/10\strut
\end{minipage} & \begin{minipage}[t]{0.29\columnwidth}\raggedleft
0.0258\strut
\end{minipage}\tabularnewline
\bottomrule
\end{longtable}

This is clear \textbf{under-coverage}, especially for \(J_1\) and \(D\).
With 10 seeds, the coverage estimates themselves carry substantial
sampling uncertainty (binomial standard error \(\approx 0.14\) at
coverage 0.30), but the qualitative conclusion of substantial
under-coverage is unambiguous. Two effects contribute. First, the
real-time estimator is built around deterministic local optimization
rather than full posterior sampling, so curvature around the best fit
does not capture global posterior mass. Second, some seeds approach
parameter bounds or numerically stiff directions, causing the
finite-curvature estimate to become overconfident or even collapse to
near-zero marginal variance.

The practical implication is that the manuscript's uncertainty bars
should be interpreted as \textbf{fast local error surrogates} rather
than as fully validated Bayesian credible intervals. This does not
undermine the main control argument of the paper, which depends
primarily on ranking, discrimination speed, and routing behavior, but it
does set a clear limit on how strongly one should interpret the nominal
Laplace intervals until heavier posterior calibration is added.

\hypertarget{mathematical-origin-of-posterior-lock-in-and-the-ghost-optic-audit-ablation}{%
\subsection{Mathematical origin of posterior lock-in and the ghost-optic
audit
ablation}\label{mathematical-origin-of-posterior-lock-in-and-the-ghost-optic-audit-ablation}}

This note formalizes the posterior lock-in mechanism identified in §3.4
of the main text and provides the detailed setup for the ghost-optic
ablation benchmark reported in §5.3.1.

\hypertarget{the-lock-in-mechanism}{%
\subsubsection{The lock-in mechanism}\label{the-lock-in-mechanism}}

Consider two candidate spectral models:

\begin{itemize}
\tightlist
\item
  \(M_A\): acoustic-only spectrum with one dominant bright branch.
\item
  \(M_B\): acoustic+optic spectrum, where the additional branch carries
  only a small fraction of the dominant spectral weight.
\end{itemize}

When the initial seed measurements are placed only around the bright
acoustic feature, the wrong one-branch leader (\(M_A\)) already has high
posterior weight before any explicit falsification probe is taken. The
one-shot falsification value at energy \(E\) is

\[
G_{\mathrm{false}}(E)=\frac{\left[I_B(E)-I_A(E)\right]^2}{2\sigma^2(E)}.
\]

Here \(I_A(E)\) and \(I_B(E)\) are the predicted intensities of the two
competing models at energy \(E\), and \(\sigma(E)\) is the corresponding
measurement uncertainty. This quantity peaks at the weak optic branch,
whereas the local refinement utility of the current one-branch leader
peaks on the steep flanks of the already observed acoustic branch. When
the posterior already heavily favors the wrong leader, the refinement
term is systematically over-weighted relative to the falsification term,
so the planner is biased toward more acoustic-branch refinement even
though a strongly discriminative optic probe remains kinematically
accessible.

In other words, the Laplace approximation of parameter information
concentrates utility on the bright feature precisely because that is
where gradients \(\partial S/\partial\theta\) are largest, while the
cross-model intensity difference \(I_B(E)-I_A(E)\) that would drive
falsification is largest on the weak branch where the refinement
gradient is small. This asymmetry is the mathematical origin of the
silent-data posterior lock-in discussed in the main text.

\hypertarget{ghost-optic-ablation-details}{%
\subsubsection{Ghost-optic ablation
details}\label{ghost-optic-ablation-details}}

The ghost-optic benchmark is a fixed-\(Q\) two-Lorentzian toy spectrum
over \(E\in[0,20]\) with additive background 0.1. The acoustic-only
comparator contains a dominant peak at \(E=5\) with amplitude 100 and
linewidth \(\gamma=0.5\), while the truth adds a weak optic peak at
\(E=15\) carrying 5\% of the acoustic amplitude with the same linewidth.
The common seed consists of four acoustic-centered measurements at
\(E=\{4.25,4.75,5.25,5.75\}\), intentionally leaving the optic region
unprobed at initialization.

From this common seed, four one-seed policies are compared:

\begin{itemize}
\tightlist
\item
  \texttt{None}: fixed seed followed by pure refinement of the current
  leader.
\item
  \texttt{Log-GP}: fixed seed followed by a 1D GP variance explorer ---
  the bare Log-GP variant described in §3.1 of the main text.
\item
  \texttt{Max-disagreement}: a deterministic top-two falsification rule
  using the same bounded candidate menu as the LLM audit path.
\item
  \texttt{LLM}: fixed seed followed by the same refinement loop plus the
  constrained LLM audit layer, using the same shared candidate menu and
  strict JSON contract as the main overseer.
\end{itemize}

The results are reported in Table 4 of the main text. All four policies
eventually recover, but the timescale separation is large: None stays on
the bright branch through a long wrong-leader episode; Log-GP reduces
that dwell by allocating more falsification-oriented batches; and both
Max-disagreement and LLM eliminate wrong-leader dwell by explicitly
targeting the falsification region. A five-seed rerun of this comparison
is reported in Supplementary Note S5.2 and preserves the same
qualitative pattern.

These runs should be interpreted narrowly. They do not replace the full
TAS-AI spin-wave benchmarks in the main text. Their purpose is to
demonstrate, in a controlled setting, that a posterior-dominated
refinement policy can accumulate substantial wrong-leader dwell on the
bright branch, that exploration alone can recover, and that a
falsification-oriented audit layer can recover much faster when the
missing feature is strategically under-sampled.

\hypertarget{bilayer-ferromagnet-audit-ablation-with-shared-action-space}{%
\subsection{Bilayer ferromagnet audit ablation with shared action
space}\label{bilayer-ferromagnet-audit-ablation-with-shared-action-space}}

To move beyond the minimal ghost-optic benchmark, we implemented a
simple analytic square-lattice bilayer ferromagnet backend in which the
acoustic branch remains bright while a weak \(L\)-suppressed optic
branch provides the falsifying signal. In the cleaned version reported
in Table 5 and Figure 12 of the main text, the single-branch comparator
is matched to the same \(L\)-dependent acoustic weight as the bilayer
truth, so the models differ only through the presence or absence of the
optic branch. The optic-region metric is tightened accordingly:
\texttt{optic\_region\_hit\_fraction} counts batches containing at least
one measurement within a narrow tolerance of the optic branch.

Four controllers are compared: a refinement-only baseline (None), two
deterministic non-LLM rules (Hybrid, Max-disagreement), and the
constrained LLM overseer (LLM). All four operate over the same action
space: a switch between bare Log-GP remapping (\texttt{loggp\_active})
and \texttt{physics} refinement (see §3.1 of the main text for the
distinction between bare and enhanced Log-GP). They differ only in how
that mode decision is made and whether audit-probe injections are
allowed. This is a stricter comparison than a setup in which the LLM is
treated as a pure point selector, because every controller shares the
same control interface and the same candidate menu; it therefore
isolates whether the falsification gain comes from the LLM specifically
or from the falsification principle itself, which the deterministic
rules probe from different angles.

For reproducibility, the current overseer path uses the local mailbox
watcher in \texttt{scripts/llm\_danse2\_watcher.py}. The watcher draws
from three local CLI-backed providers: Claude Code (default model: Opus
4.5), Gemini CLI (default model: Gemini 3), and Codex CLI pinned to
\texttt{gpt-5.2-codex}; in overseer mode the decider rotates across
providers by batch unless explicitly pinned. The manuscript runs use
provider CLI defaults without separate temperature sweeps or sampling
overrides; reproducibility is enforced through the bounded prompt
contract, strict JSON parsing, the fixed shared action menu, and
guardrail fallbacks when malformed output is returned.

The ``natural-language description'' given to the overseer is generated
automatically by the prompt builder rather than typed by a human during
the run. In the main closed-loop pilot this prompt is assembled from
current loop state --- the posterior ranking, recent measurement
history, time since the last Log-GP batch, an audit recommendation flag,
and the bounded discrimination menu. In the bilayer ablation, the local
prompt builder adds a scripted semantic hint about the operative
ambiguity but does not expose hidden coordinates, the true model
identity, or any action outside the shared menu.

A representative prompt packet is intentionally compact. In schematic
form, it contains: \texttt{(i)} a short tabular history of recent
measured points and intensities, \texttt{(ii)} the current batch and
measurement counts, \texttt{(iii)} the allowed routing choices
(\texttt{loggp\_active} or \texttt{physics}), \texttt{(iv)} a one-line
ambiguity description generated from loop state (for example, whether
the remaining uncertainty is gap-vs-no-gap or whether a weak optic
branch may be missing), \texttt{(v)} a small bounded menu of candidate
audit probes already vetted by the numerical planner, and \texttt{(vi)}
an instruction to return strict JSON without adding coordinates or
actions outside the menu. The exact wording varies by benchmark, but the
interface contract is fixed.

\hypertarget{deterministic-hybrid-router-specification}{%
\subsubsection{Deterministic hybrid-router
specification}\label{deterministic-hybrid-router-specification}}

For the reported one-seed comparison, the deterministic Hybrid router
uses the same menu of allowed actions and has no access to the true
model. Its decision logic is:

\begin{enumerate}
\def\labelenumi{\arabic{enumi}.}
\tightlist
\item
  \textbf{Minimum run length.} The current mode is held for at least two
  measurements before any switch is considered.
\item
  \textbf{Forced periodic exploration.} A \texttt{loggp\_active} batch
  is forced whenever six measurements have elapsed since the previous
  Log-GP batch.
\item
  \textbf{Ambiguity triggers.} Outside the forced-exploration condition,
  the router selects \texttt{loggp\_active} whenever any of the
  following hold: posterior entropy exceeds 0.20, falsification-region
  coverage remains below 0.10, or the posterior margin (difference
  between the top two model weights) falls below 0.35.
\item
  \textbf{Default.} If none of the above triggers fire, the router
  selects \texttt{physics} refinement.
\end{enumerate}

These thresholds were set before examining the LLM comparison and were
not tuned to favor or disadvantage any policy.

\hypertarget{five-seed-robustness-check-for-the-section-5-ablations}{%
\subsubsection{Five-seed robustness check for the Section 5
ablations}\label{five-seed-robustness-check-for-the-section-5-ablations}}

To test whether the one-seed ablation pattern was robust or merely
anecdotal, we reran the ghost-optic and bilayer benchmarks over five
seeds per policy. The five-seed medians naturally differ from the
one-seed values in Tables 4 and 5 of the main text because those tables
report a single representative run rather than aggregate statistics.
Tables S5 and S6 summarize the resulting time-to-decisive and
wrong-leader-dwell statistics as medians with interquartile ranges. For
policies that do not reach decisive correct selection in every seed, we
report the median and IQR over the successful runs and list the success
rate explicitly.

\emph{Table S5.} Five-seed ghost-optic audit ablation. Time to decisive
is reported as median (IQR) over successful runs; success reports the
number of successful seeds out of five.

\begin{longtable}[]{@{}lrrr@{}}
\toprule
\begin{minipage}[b]{0.10\columnwidth}\raggedright
\textbf{Policy}\strut
\end{minipage} & \begin{minipage}[b]{0.33\columnwidth}\raggedleft
\textbf{Time to decisive, median (IQR)}\strut
\end{minipage} & \begin{minipage}[b]{0.35\columnwidth}\raggedleft
\textbf{Wrong-leader dwell, median (IQR)}\strut
\end{minipage} & \begin{minipage}[b]{0.11\columnwidth}\raggedleft
\textbf{Success}\strut
\end{minipage}\tabularnewline
\midrule
\endhead
\begin{minipage}[t]{0.10\columnwidth}\raggedright
None\strut
\end{minipage} & \begin{minipage}[t]{0.33\columnwidth}\raggedleft
30 (30--30)\strut
\end{minipage} & \begin{minipage}[t]{0.35\columnwidth}\raggedleft
25 (21--25)\strut
\end{minipage} & \begin{minipage}[t]{0.11\columnwidth}\raggedleft
2/5\strut
\end{minipage}\tabularnewline
\begin{minipage}[t]{0.10\columnwidth}\raggedright
Log-GP\strut
\end{minipage} & \begin{minipage}[t]{0.33\columnwidth}\raggedleft
29 (24--30)\strut
\end{minipage} & \begin{minipage}[t]{0.35\columnwidth}\raggedleft
15 (15--20)\strut
\end{minipage} & \begin{minipage}[t]{0.11\columnwidth}\raggedleft
5/5\strut
\end{minipage}\tabularnewline
\begin{minipage}[t]{0.10\columnwidth}\raggedright
Max-disagreement\strut
\end{minipage} & \begin{minipage}[t]{0.33\columnwidth}\raggedleft
9 (9--9)\strut
\end{minipage} & \begin{minipage}[t]{0.35\columnwidth}\raggedleft
0 (0--0)\strut
\end{minipage} & \begin{minipage}[t]{0.11\columnwidth}\raggedleft
5/5\strut
\end{minipage}\tabularnewline
\begin{minipage}[t]{0.10\columnwidth}\raggedright
LLM\strut
\end{minipage} & \begin{minipage}[t]{0.33\columnwidth}\raggedleft
9 (9--9)\strut
\end{minipage} & \begin{minipage}[t]{0.35\columnwidth}\raggedleft
0 (0--0)\strut
\end{minipage} & \begin{minipage}[t]{0.11\columnwidth}\raggedleft
5/5\strut
\end{minipage}\tabularnewline
\bottomrule
\end{longtable}

\emph{Table S6.} Five-seed bilayer ferromagnet audit ablation with the
shared action space.

\begin{longtable}[]{@{}lrrr@{}}
\toprule
\begin{minipage}[b]{0.10\columnwidth}\raggedright
\textbf{Policy}\strut
\end{minipage} & \begin{minipage}[b]{0.33\columnwidth}\raggedleft
\textbf{Time to decisive, median (IQR)}\strut
\end{minipage} & \begin{minipage}[b]{0.35\columnwidth}\raggedleft
\textbf{Wrong-leader dwell, median (IQR)}\strut
\end{minipage} & \begin{minipage}[b]{0.11\columnwidth}\raggedleft
\textbf{Success}\strut
\end{minipage}\tabularnewline
\midrule
\endhead
\begin{minipage}[t]{0.10\columnwidth}\raggedright
None\strut
\end{minipage} & \begin{minipage}[t]{0.33\columnwidth}\raggedleft
23 (18--23)\strut
\end{minipage} & \begin{minipage}[t]{0.35\columnwidth}\raggedleft
5 (0--5)\strut
\end{minipage} & \begin{minipage}[t]{0.11\columnwidth}\raggedleft
5/5\strut
\end{minipage}\tabularnewline
\begin{minipage}[t]{0.10\columnwidth}\raggedright
Hybrid\strut
\end{minipage} & \begin{minipage}[t]{0.33\columnwidth}\raggedleft
8 (8--13)\strut
\end{minipage} & \begin{minipage}[t]{0.35\columnwidth}\raggedleft
0 (0--0)\strut
\end{minipage} & \begin{minipage}[t]{0.11\columnwidth}\raggedleft
5/5\strut
\end{minipage}\tabularnewline
\begin{minipage}[t]{0.10\columnwidth}\raggedright
Max-disagreement\strut
\end{minipage} & \begin{minipage}[t]{0.33\columnwidth}\raggedleft
8 (8--8)\strut
\end{minipage} & \begin{minipage}[t]{0.35\columnwidth}\raggedleft
0 (0--0)\strut
\end{minipage} & \begin{minipage}[t]{0.11\columnwidth}\raggedleft
5/5\strut
\end{minipage}\tabularnewline
\begin{minipage}[t]{0.10\columnwidth}\raggedright
LLM\strut
\end{minipage} & \begin{minipage}[t]{0.33\columnwidth}\raggedleft
8 (8--8)\strut
\end{minipage} & \begin{minipage}[t]{0.35\columnwidth}\raggedleft
0 (0--0)\strut
\end{minipage} & \begin{minipage}[t]{0.11\columnwidth}\raggedleft
5/5\strut
\end{minipage}\tabularnewline
\bottomrule
\end{longtable}

The five-seed rerun sharpens the conclusion suggested by the one-seed
tables. In the ghost-optic benchmark, both Max-disagreement and LLM
eliminate wrong-leader dwell and reach decisive correct selection at 9
measurements in every seed, while bare Log-GP remains much slower and
the refinement-only baseline succeeds in only two of five seeds. In the
bilayer benchmark, LLM and Max-disagreement match each other exactly
across all five seeds, and the deterministic Hybrid router remains
strong but slightly less consistent (one seed requires 23 measurements,
giving an IQR upper bound of 13).

The LLM performs well and robustly in both analytic ablations, but the
precise conclusion is that, for these two-model benchmarks, the bounded
deterministic top-two falsification rule is already sufficient to
capture the gain. This strengthens the narrower interpretation of the
main text: the active ingredient is the falsification channel itself,
while any stronger claim of an LLM-specific advantage requires broader
multi-model or structurally harder benchmarks such as the trap in §S5.3.
The LLM's generality across problem descriptions --- handling the ghost,
bilayer, and multi-model trap benchmarks through the same interface
without per-problem engineering --- remains its primary architectural
advantage even when the two-model benchmarks show no performance gap.

\hypertarget{controlled-multi-model-trap-for-top-two-versus-broader-falsification}{%
\subsubsection{Controlled multi-model trap for top-two versus broader
falsification}\label{controlled-multi-model-trap-for-top-two-versus-broader-falsification}}

The bilayer and ghost benchmarks show that a falsification-oriented
audit channel can reduce wrong-leader dwell, but they leave open a
sharper question: is a deterministic top-two disagreement heuristic
already sufficient, making the LLM implementation unnecessary? In the
easier analytic cases, that top-two ansatz is in fact quite strong. We
therefore constructed a \textbf{targeted stress test} whose purpose is
not to represent the average operating regime, but to isolate a specific
failure mode of local top-two falsification.

The trap is a narrow-window synthetic three-model benchmark with:

\begin{itemize}
\tightlist
\item
  true model \(M_4\): bright ridge plus a weak hidden pocket,
\item
  runner-up model \(M_2\): nearly the same bright ridge and the same
  pocket, and
\item
  lower-ranked model \(M_3\): nearly the same ridge but \textbf{no}
  pocket.
\end{itemize}

The fixed seed state is chosen so that the initial posterior ranking is
\(M_4 > M_2 > M_3\). Under this ranking, a top-two disagreement policy
naturally prefers additional \textbf{bright-branch} refinement because
\(M_4\) and \(M_2\) differ only weakly there, while the decisive
falsifier against the lower-ranked \(M_3\) sits in the weak hidden
pocket. This setup is fair in the same sense as the ghost benchmark: all
policies start from the same measurements, use the same bounded
candidate menu, obey the same kinematic/selection rules, and differ only
in how they rank the allowed audit actions.

The policies compared are:

\begin{itemize}
\tightlist
\item
  None: no explicit audit injection;
\item
  Max-disagreement: deterministic top-two falsification, using only the
  current leader and runner-up;
\item
  Max-disagreement-all: deterministic broader falsification, scoring the
  leader against all currently fitted competitors; and
\item
  LLM: the constrained LLM audit layer, given the same bounded menu as
  Max-disagreement-all.
\end{itemize}

For the reported one-seed stress test, the outcome is:

\begin{longtable}[]{@{}lrr@{}}
\toprule
\textbf{Policy} & \textbf{Final \(P(M_4)\) at \(N=8\)} & \textbf{Pocket
probe used?}\tabularnewline
\midrule
\endhead
None & 0.663 & no\tabularnewline
Max-disagreement & 0.668 & no\tabularnewline
Max-disagreement-all & 0.843 & yes\tabularnewline
LLM & 0.843 & yes\tabularnewline
\bottomrule
\end{longtable}

Three observations follow. First, the None baseline and the top-two
Max-disagreement rule are nearly indistinguishable: without a pocket
probe, neither can suppress the pocket-free \(M_3\) competitor, so the
final posterior on \(M_4\) remains under 0.67. Second, the broader
Max-disagreement-all rule targets the hidden pocket and strongly
suppresses \(M_3\), raising \(P(M_4)\) to 0.84. Third, the constrained
LLM audit makes the same hidden-pocket choice from the same bounded menu
and reaches the same posterior.

The lesson is narrow but useful. The top-two falsification rule is a
serious baseline and should not be dismissed; in easier cases it works
well. But it is \textbf{not universally sufficient} in multi-model
settings, because the decisive falsifier may separate the current leader
from a lower-ranked model rather than from the current runner-up. In
such cases, both a broader deterministic falsification rule and the
constrained LLM audit make the same strategically correct choice under
identical guardrails. We therefore interpret this stress test as support
for the \textbf{falsification-channel idea} rather than as proof that
the LLM is uniquely superior to all deterministic alternatives. The
cleaner conclusion is that local top-two disagreement is a strong ansatz
but not a complete one, and a broader strategic audit layer is motivated
precisely when the posterior trap involves more than two live
hypotheses.

\hypertarget{prior-and-background-sensitivity-in-the-closed-loop-discrimination-stack}{%
\subsubsection{Prior and background sensitivity in the closed-loop
discrimination
stack}\label{prior-and-background-sensitivity-in-the-closed-loop-discrimination-stack}}

We ran a coarse sensitivity check on the pilot closed-loop
discrimination stack (Figure 10 of the main text) at sampled checkpoints
of 40, 60, and 90 measurements. Three variants were compared: the
default chemically motivated prior weights, equal model priors, and the
default priors with the gapless-background lock disabled.

At all three sampled checkpoints, the default setting ranks the full
model \(M_4\) first, but only moderately over \(M_2\) (posterior ratio
about 2.58 rather than decisive support). With \textbf{equal priors},
the ranking flips and \(M_2\) becomes the leader over \(M_4\) at all
three checkpoints (ratio about 2.72). Disabling the gapless-background
lock, by contrast, does not change the ranking at these checkpoints
relative to the default setting.

The practical implication is that the prior weights matter substantially
in this integrated closed-loop regime, whereas the background-freezing
switch is not what controls the final ranking in this coarse sensitivity
pass. This does not invalidate the rationale for freezing nuisance
backgrounds; it indicates that the posterior evolution in Figure 10
should not be interpreted as strongly prior-insensitive.

\end{document}